%% file: main.tex
\documentclass[sigconf]{acmart} 
\acmConference[ICSE 2026]{the 48th International Conference on Software Engineering}{April 12--18,
  2026}{Rio de Janeiro, Brazil}

\usepackage{graphicx}
\usepackage{xcolor}
\usepackage{booktabs}
\usepackage{enumerate}
\usepackage{hyperref}
\usepackage{url}
\usepackage{amsthm}
\usepackage{makecell}
\usepackage{multirow}
\usepackage[normalem]{ulem} 
\newcommand{\ts}{\bgroup\markoverwith {\textcolor{red2}{\rule[0.5ex]{2pt}{0.8pt}}}\ULon} 
\hypersetup{
    colorlinks=false, 
    pdfborder={0 0 0}, 
    pdfpagemode=UseNone, 
    hidelinks 
}

\usepackage{caption, subcaption}
\usepackage{listings}
\input{listing_pls/listing_python}

\usepackage[linesnumbered,ruled,vlined]{algorithm2e}

\usepackage{enumitem}
\setlist[enumerate]{leftmargin=1.5em}
\setlist[itemize]{leftmargin=0.8em}

\newtheoremstyle{tight}  
  {1.5pt}   
  {1.5pt}   
  {}      
  {}      
  {\bfseries} 
  {.}     
  { }     
  {}      

\theoremstyle{tight} 
\newtheorem{definition}{Definition}

\newcommand{\code}[1]{\texttt{\small#1}}

\newcommand{\name}{Code\-Mapper}

\usepackage{tcolorbox}
\newcounter{findingCounter}
\newenvironment{finding}{
\begin{tcolorbox}[colback=blue!5!white,colframe=blue!5!white,arc=0mm,grow to left by=0mm,left=0mm,grow to right by=0mm,left=1.2mm,right=1.2mm,top=1.2mm,bottom=1.2mm]
\textbf{Answer to RQ\arabic{findingCounter}\stepcounter{findingCounter}:}}
{
\end{tcolorbox}
}

\copyrightyear{2026}
\acmYear{2026}
\setcopyright{cc}
\setcctype{by}
\acmConference[ICSE '26]{2026 IEEE/ACM 48th International Conference on Software Engineering}{April 12--18, 2026}{Rio de Janeiro, Brazil}
\acmBooktitle{2026 IEEE/ACM 48th International Conference on Software Engineering (ICSE '26), April 12--18, 2026, Rio de Janeiro, Brazil}
\acmPrice{}
\acmDOI{10.1145/3744916.3773267}
\acmISBN{979-8-4007-2025-3/2026/04}

\begin{document}

\title[\name{}: A Language-Agnostic Approach to Mapping Code Regions Across Commits]{\name{}: A Language-Agnostic Approach to\\ Mapping Code Regions Across Commits}

\author{Huimin Hu}
\email{huhuimin236@gmail.com}
\orcid{0000-0002-1470-0839}
\affiliation{%
  \institution{CISPA Helmholtz Center for Information Security}
  \city{Stuttgart}
  \country{Germany}
}

\author{Michael Pradel}
\email{michael@binaervarianz.de}
\orcid{0000-0003-1623-498X}
\affiliation{%
  \institution{CISPA Helmholtz Center for Information Security}
  \city{Stuttgart}
  \country{Germany}
}

\begin{abstract}
During software evolution, developers commonly face the problem of mapping a specific code region from one commit to another.
For example, they may want to determine how the condition of an if-statement, a specific line in a configuration file, or the definition of a function changes.
We call this the \emph{code mapping problem}.
Existing techniques, such as git diff, address this problem only insufficiently because they show all changes made to a file instead of focusing on a code region of the developer's choice.
Other techniques focus on specific code elements and programming languages (e.g., methods in Java), limiting their applicability.
This paper introduces \name{}, an approach to address the code mapping problem in a way that is independent of specific program elements and programming languages.
Given a code region in one commit, \name{} finds the corresponding region in another commit.
The approach consists of two phases: (i) computing candidate regions by analyzing diffs, detecting code movements, and searching for specific code fragments, and (ii) selecting the most likely target region by calculating similarities.
Our evaluation applies \name{} to four datasets, including two new hand-annotated datasets containing code region pairs in ten popular programming languages.
\name{} correctly identifies the expected target region in 71.0\%--94.5\% of all cases, improving over the best available baselines by 1.5--58.8 absolute percent points.
\end{abstract}

  

\maketitle

\begin{figure*}[tbp]
\centering
\begin{subfigure}[t]{0.34\textwidth}
    \centering
    \includegraphics[width=\textwidth]{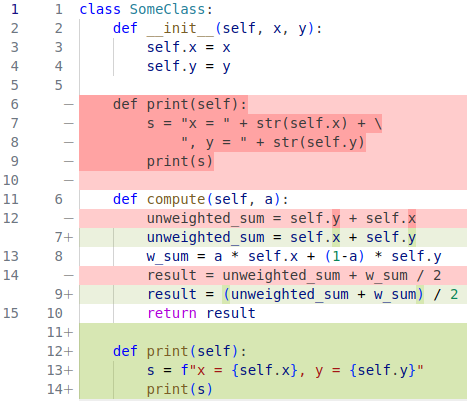}
    \captionsetup{skip=3pt}
    \caption{Line-level diff.}
    \label{subfig:motivation_diff}
\end{subfigure}\hfill%
\begin{subfigure}[t]{0.65\textwidth}
    \centering
    \raisebox{6.8\baselineskip}{%
        \begin{minipage}{\linewidth}
            \includegraphics[width=\textwidth]{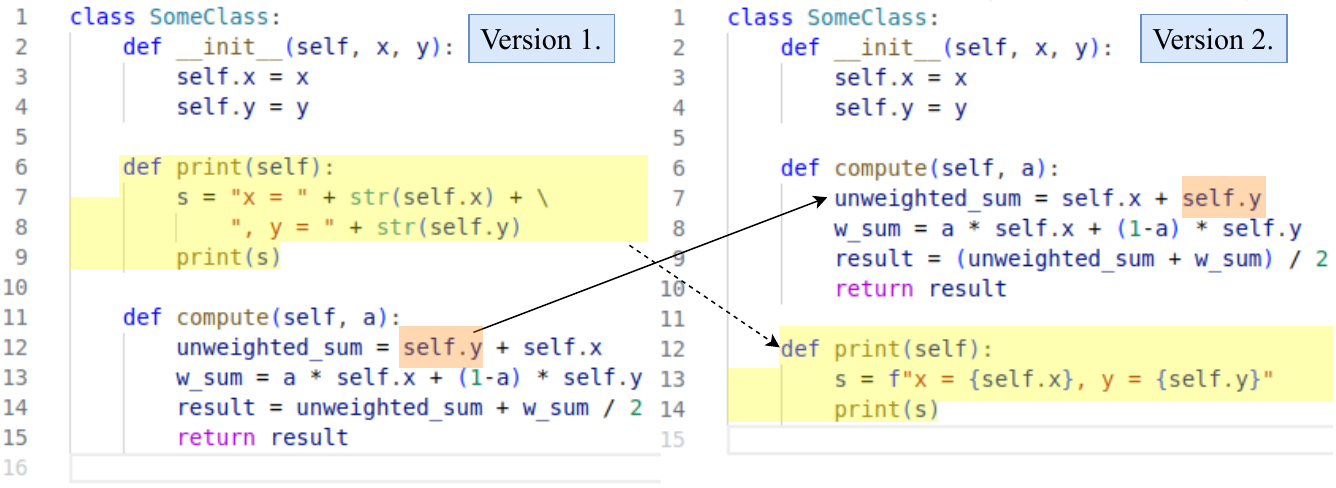}
            \captionsetup{justification=centering,skip=3pt}
            \caption{\name{} maps a larger region (highlighted in yellow)\\
            and a smaller region (highlighted in orange).}
            \label{subfig:motivation_track_regions}
        \end{minipage}
    }%
\end{subfigure}
\captionsetup{skip=5pt}
\caption{Example of mapping code regions in Python. Red and green highlighting is code that got deleted and added according to git diff. The yellow and orange highlighting shows correctly mapped code regions, as provided by \name{}.}
\label{fig:motivation_example}
\end{figure*}

\section{Introduction}
\label{sec:introduction}
Tracking code across different commits during the evolution of a project is a vital step for many software development tasks.
An empirical study~\cite{Codoban2015} reports that developers examine software histories for a variety of reasons, such as keeping up with changes done by others, understanding the impact of code that the developers are currently developing, and identifying changes that have introduced an error.
The study highlights that developers are often most interested in changes affecting their current task, and that a common strategy is to traverse commits with a specific goal in mind.
These observations imply that developers do not necessarily want to see all changes, but rather map specific code regions of interest from one commit to another.
Questions on Stack Overflow also show that developers want to identify and track specific changes, 
e.g., by checking the evolution of a variable,\footnote{\href{https://stackoverflow.com/questions/9935379/git-show-all-of-the-various-changes-to-a-single-line-in-a-specified-file-over-t}{How to track changes to a variable's value in a JavaScript file?~\cite{post1}} and 
  \mbox{\href{https://stackoverflow.com/questions/34576699/git-grep-file-in-all-previous-versions?noredirect=1}{How to track changes to a specific value in a Python config file?~\cite{post2}}}
}
or by finding changes on a specific code region.\footnote{\href{https://stackoverflow.com/questions/40936797/how-can-i-use-git-log-and-only-output-the-matching-lines?rq=3}{Can git diff show only the lines around a specific term?~\cite{post3}}}
The discussions around these posts show that currently available tools do not fully address developers' needs.

Git, a popular version control system, provides git diff, with various options to show code changes.
By default, git diff computes diffs at the line-level, i.e., showing which lines are removed and added. 
It also supports word-level diffs with the ``--word-diff'' option.
While git diff is powerful, it has limitations when it comes to mapping specific code regions from one commit to another.
One limitation is that it reports all changes. 
However, as noted by Codoban et al.~\cite{Codoban2015} and the posts mentioned above, developers sometimes prefer to focus on specific parts of a change, and they struggle to find a tool for this purpose.
Another limitation is that it may fail to accurately match code fragments, especially when the changes are complex. 
Git also provides ``git log -L'', which shows the commit history for a specified range of lines.
However, it relies on Git's history tracking mechanisms, which also underlie git diff, and hence suffers from the same limitations.

Figure~\ref{fig:motivation_example} shows an example demonstrating the limitations of git diff.
From version~1 to version~2, the function \code{print} is modified and relocated. However, git diff identifies the changes as deleting code and adding new code  (Figure~\ref{subfig:motivation_diff}). 
Instead, a developer interested in mapping this function from version~1 to version~2 would benefit from a tool that accurately recognizes the movement (Figure~\ref{subfig:motivation_track_regions}, yellow).
The git diff output also makes it difficult to see how the usage of the \code{self.y} attribute in the \code{compute} function evolves, which ideally should be shown as illustrated in Figure~\ref{subfig:motivation_track_regions} (orange).

We refer to the problem of mapping a specific code region from one commit of a project to another commit as the \emph{code mapping problem}.
To address this problem, we present \name{}, a code mapping approach designed to focus on specific code regions of the developer's choice. 
Given a code region in one commit, \name{} finds the corresponding region in another commit.
The approach works in two phases. 
In the first phase, it computes candidate regions by running several techniques with complementary strengths. 
Specifically, the first phase runs git diff with several algorithms and levels of granularity to identify hunks that modify or relocate the source region. 
The approach then checks for any moved code and provides details if a region got relocated. 
Additionally, the approach performs a text search to find exact occurrences of the given code region in the new file. 
In the second phase, \name{} selects the most likely candidate region by computing the similarity between the source region and all candidate regions, while also considering the code around these regions.

Our work on addressing the code mapping problem is not meant to replace git diff, but rather to complement it by enabling developers to focus on a specific code region of their choice.
While prior work has addressed related problems, we are not aware of any existing approach that addresses the general code mapping problem.
One related line of work focuses on tracking specific code elements, such as a method or a variable, across the entire version history of a project~\cite{icse_2021_codeshovel,10.1145/3540250.3549079,Hasan:TSE:2024:CodeTracker2.0}.
That work gets evaluated based on its ability to find the commits that modify the specific code element and to identify the kinds of changes made to the code element in these commits.
In contrast, our problem formulation assumes two commits to be given and focuses on mapping a specific code region from one commit to the other.
Another line of work creates a new version history in which each method is stored in a separate file named by its fully qualified class and method name~\cite{2011Historage,2020FinerGit}.
In contrast, our approach does not require rewriting history, and it supports tracking arbitrary code regions, instead of focusing on specific code elements.
\name{} also differs from all the above work by being language-agnostic, instead of relying on language-specific parsers and heuristics, making it more widely applicable.



We evaluate \name{} on four datasets: 
two newly created datasets with annotated code regions of various sizes and in ten popular languages, 
an existing dataset containing histories of Python comments that suppress static analysis warnings, and a dataset derived from prior work on tracking the history of code elements in Java~\cite{10.1145/3540250.3549079,Hasan:TSE:2024:CodeTracker2.0}.
The results demonstrate \name{}'s effectiveness in mapping code regions and clear improvements over tools that are currently used in practice (line-level and word-level git diff). 
Depending on the dataset, the approach achieves an exact match rate of 71.0\%--94.5\%, with a recall of 78.1\%--97.7\% and a precision of 76.4\%--97.4\%.
The results improve over the best available baselines, e.g., by 1.5--58.8 absolute percent points in terms of exact match rate.
In addition to being more effective, \name{} is also sufficiently efficient for interactive usage, with an average execution time of 2,327 milliseconds to map a code region.

In summary, this work makes the following contributions:
\begin{itemize}
  \item The first approach to the code mapping problem that is independent of a specific programming language and the kind of code region to map.  
  \item Two reusable datasets, together containing 200 carefully annotated pairs of code regions mapped across pairs of commits in ten popular programming languages.  
  \item Experiments that demonstrate \name{} to clearly outperform the current state of the art (git diff) in terms of effectiveness, while providing an efficiency that is on par with currently used tools.
\end{itemize}

\section{Approach}


\subsection{Terminology and Problem Definition}
\label{sec:terminology}


Before presenting our approach, we define important terms used in this paper and the problem we are addressing.
When developers change code, they transform an old version of the code into a new version.
Because \name{} supports mapping code regions both forward and backward in time, we avoid using terms like ``old'' and ``new'' to refer to versions.
Instead, we refer to the version from where the region gets mapped as the \emph{source} and the version where the region gets mapped to as the \emph{target}. 
To identify a contiguous block of changed code, we use the common term \emph{hunk}, i.e., a contiguous block of lines that have been added, deleted, or modified between two file versions:
\begin{definition}[Hunk]
  \label{def:hunk}
  A hunk is a tuple,
  $H = (\mathit{l}_{\mathit{source}}^{\mathit{start}}, l_{\mathit{source}}^{\mathit{end}},\\l_{\mathit{target}}^{\mathit{start}}, l_{\mathit{target}}^\mathit{end})$,
  where $l_{\mathit{source}}^{\mathit{start}}$ and $l_{\mathit{source}}^{\mathit{end}}$ are the first and last line of the changed block in the source version,
  and $l_{\mathit{target}}^{\mathit{start}}$ and $l_{\mathit{target}}^{\mathit{end}}$ are the first and the last line of the changed block in the target version.
\end{definition}

To define a fragment of code to map, we use character-level granularity:
\begin{definition}[Character range]
    \label{def:character_range}
    A character range is a tuple, 
    \textit{R = ($\mathit{l}_{\mathit{1}}$, $\mathit{c}_{\mathit{1}}$, $\mathit{l}_{\mathit{2}}$, $\mathit{c}_{\mathit{2}}$)}, where
    $\mathit{l}_{\mathit{1}}$ and $\mathit{c}_{\mathit{1}}$ are the line number and the character number where the range starts,
    and $\mathit{l}_{\mathit{2}}$ and $\mathit{c}_{\mathit{2}}$ are the line number and the character number where the range ends.
\end{definition}

All numbers in such a tuple start at one, $\mathit{l}_{\mathit{1}} <= \mathit{l}_{\mathit{2}}$, and a range includes at least one character.
Based on the character range, we define a region of code:
\begin{definition}[Region]
    \label{def:region}
    A region is a tuple, \textit{G = (c, f, R)} where $c$ is a commit hash, $f$ is the file path of the file containing the region, and $R$ is the character range of the region.
\end{definition}

Finally, we define the problem we are addressing, which is to map a given region from one commit to its corresponding region in another commit:
\begin{definition}[Code mapping problem]
Given a source region $G_{\mathit{source}} = (c_{\mathit{source}}, f_{\mathit{source}}, R_{\mathit{source}})$ and a target commit $c_{\mathit{target}}$, 
determine the corresponding target region $\mathit{G}_{\mathit{target}} = (c_{\mathit{target}}, f_{\mathit{target}}, R_{\mathit{target})}$.
\end{definition}
The target region can fall into one of three cases:
First, the source region and target region are the same, because the code in $f_{\mathit{source}}$ has not changed.
Second, the target region differs from the source region.
Third, the source region does not exist anymore in the target commit, e.g., because the code was deleted, which we represent with the special value $G_{\mathit{target}}=(\perp, \perp, \emptyset)$.
Our work supports all three of these cases.
Moreover, we address this problem without assuming a specific programming language or kind of program element to map, allowing developers to map arbitrary code regions across different versions of a project.

Note that the code mapping problem differs from the code tracking problem~\cite{icse_2021_codeshovel,10.1145/3540250.3549079,Hasan:TSE:2024:CodeTracker2.0} in terms of what is assumed to be given, as well as when and how developers may want to use approaches that address these problems.
For code tracking, a single commit $c$ and a specific code element $e$ (e.g., a variable or a method) are given, and the task is to find all commits that modify $e$ and the kinds of changes made to $e$ in these commits.
This is useful when developers want to understand the history of a specific code element, e.g., to understand how a method evolved over the project's lifetime.
In contrast, the code mapping problem, as addressed here, assumes two commits $c_{\mathit{source}}$ and $c_{\mathit{target}}$, and a specific code region in the source commit, to be given.
The code region may correspond to a specific code element, but it may also be a fragment of a code element or encompass multiple code elements.
The code mapping problem is useful when developers want to understand how the differences between two commits impact a specific code region, e.g., when reviewing commits made by others.

\subsection{Overview}

\begin{figure}[t]
  \centering
  \includegraphics[width=0.46\textwidth]{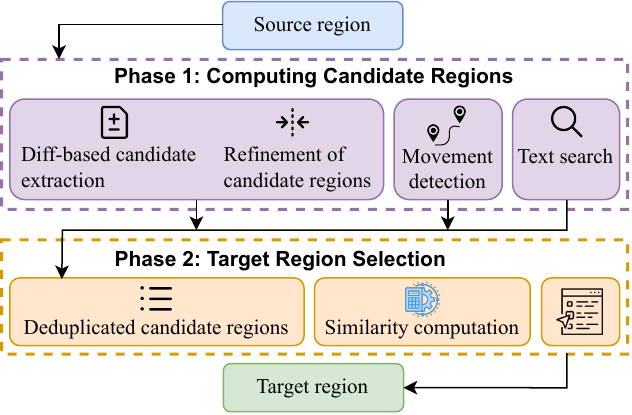}
  \vspace{-0.1in}
  \caption{Overview of \name{}.}
  \label{fig:overview}
\end{figure}

Figure~\ref{fig:overview} gives an overview of our approach for tackling the mapping problem.
The approach receives a source region and yields the corresponding target region.
It consists of two phases: computing candidate regions and selecting the target region. 

In phase 1, \name{} uses three techniques to compute candidates for the target region.
The motivation for using multiple techniques is that no single technique is perfect on its own.
Instead, by combining these techniques, we increase the ability of \name{} to identify the correct region.
The first technique builds on a standard diff computation to obtain hunks, extracts those hunks that either modify the source region or affect its location, identifies candidate regions from those hunks, and then further refines them to precisely identify the beginning and end of the region. 
The second technique builds on the hunks to detect moved code. 
Finally, the third technique searches for occurrences of the text in the character range of the source region in the target file, which is motivated by the fact that the hunks may be inaccurate.
Each of the three techniques produces a set of candidates, which the approach deduplicates and then gives to phase~2.

In phase 2, \name{} selects the most likely candidate as the target region by computing the similarity between each candidate region and the source region.
To compute the similarity between two regions, the approach compares the code in the regions themselves, as well as some contextual code, using Levenshtein distance. 

\SetKwInput{Input}{Input}
\SetKwInput{Output}{Output}
\begin{algorithm}[t]
  \caption{\name{}}
  \label{alg:overall_algorithm}
  \small\Input{Source region $\mathit{G}_{\mathit{source}}: (\mathit{c}_{\mathit{source}}, \mathit{f}_{\mathit{source}}, \mathit{R}_{\mathit{source}})$,
  \\ \qquad \quad repository $\mathit{repo}$, target commit $\mathit{c}_{\mathit{target}}$}
  \small\Output{Target region $\mathit{G}_{\mathit{target}}$: $(\mathit{c}_{\mathit{target}}, \mathit{f}_{\mathit{target}}, \mathit{R}_{\mathit{target}})$}
  \tcp{\small\textbf{Phase 1: Computing candidate regions}}
  $\mathit{Candidates}, \mathit{overlappingInfo}  \gets \emptyset$ \\
  $\mathit{diffs} \gets \mathit{getDiffReports}(\mathit{repo}, \mathit{c}_{\mathit{source}}, \mathit{c}_{\mathit{target}}, \mathit{f}_{\mathit{source}})$ \label{overall_algorithm:diffs_start} \\
  
  \For{$\mathit{diff}$ in $\mathit{diffs}$}{  \label{overall_algorithm:diffs_end}
    \For{$\mathit{H}$ in $\mathit{getHunks}(\mathit{diff})$}{ \label{overall_algorithm:hunk_iteration_start} 
      $\mathit{overlapLoc} \gets \mathit{checkOverlap}(\mathit{H}, \mathit{R}_{\mathit{source}})$ \\
      \If{$\mathit{overlapLoc} == \mathit{``fully \ covered}$''}{ \label{overall_algorithm:fully_cover_start} 
        $\mathit{addUniqueCandi}(Candidates,$ $\mathit{getRefinedRanges}(\mathit{H}, \mathit{f}_{\mathit{source}}, \mathit{R}_{\mathit{source}}))$ 
      } \label{overall_algorithm:fully_cover_end} 
      \ElseIf{$\mathit{overlapLoc}$ in \{``top'', ``middle'', ``bottom''\}}{ \label{overall_algorithm:overlalpping_start} 
        $\mathit{overlappingInfo} \gets (\mathit{overlapLoc}, \mathit{H})$ \\
      } \label{overall_algorithm:overlalpping_end} 
      \ElseIf{$\mathit{overlapLoc} == \mathit{``disjoint}$''}{ \label{overall_algorithm:not_related_start}
        \If{$\mathit{H}.\mathit{l}_{\mathit{source}}^{\mathit{end}} < \mathit{R}_{\mathit{source}}.l_1$}{ \label{overall_algorithm:not_related_check}
          $\mathit{updateCandidateLineNums}(\mathit{R}_{\mathit{source}}, \mathit{H})$ \label{overall_algorithm:not_related_update_linenum}
        }
      }\label{overall_algorithm:not_related_end}
      \If{$\mathit{lineNums}_{\mathit{overlapping}} == \mathit{lineNums}_{\mathit{source}}$}{ \label{overall_algorithm:break_checking}
        \textbf{break}
      }\label{overall_algorithm:break}
    } \label{overall_algorithm:hunk_iteration_end} 
  }
  \For{$\mathit{info}$ in $\mathit{overlappingInfo}$}{ \label{overall_algorithm:records_iteration_start}
    $\mathit{addUniqueCandi}(\mathit{Candidates}, \mathit{getCandidates}(\mathit{info}))$
  } \label{overall_algorithm:records_iteration_end}
  \If{$\mathit{R}_{\mathit{source}}$ is $\mathit{fully\_deleted}$}{ \label{overall_algorithm:fully_delete_checking_start}
    $\mathit{addUniqueCandi}(\mathit{Candidates}, \mathit{detectMovements}(\mathit{R}_{\mathit{source}}))$
  }\label{overall_algorithm:fully_delete_checking_end}
  $\mathit{addUniqueCandi}(\mathit{Candidates}, \mathit{searchTexts}(\mathit{G}_{\mathit{source}}))$ \label{overall_algorithm:search_candidates}\\
  \tcp{\small\textbf{Phase 2: Target region selection}}
  $\mathit{similaries} \gets \mathit{computeSimilarity}(\mathit{G}_{\mathit{source}}, \mathit{Candidates})$ \label{overall_algorithm:similarity_computation}\\
  $\mathit{targetIdx} \gets \mathit{similaries.index(\mathit{max(similarities)})}$ \label{overall_algorithm:target_idx_computation}\\
  $\mathit{G}_{\mathit{target}} \gets \mathit{Candidates[targetIdx]}$ \label{overall_algorithm:select_target}\\
  \Return{$\mathit{G}_{\mathit{target}}$} \label{overall_algorithm:return}
\end{algorithm}

Algorithm~\ref{alg:overall_algorithm} provides a more detailed summary of the two phases of our approach, with further discussion in Sections~\ref{section:candidate_computation} and~\ref{section:target_selection}.

\subsection{Computing Candidate Regions}
\label{section:candidate_computation}

\name{} employs three techniques to compute candidates for the target region, as described in the following.

\subsubsection{Diff-Based Candidate Extraction}
\label{section:diff_based_candidate_extraction}

This technique uses hunks in a diff report to extract candidate regions.
The approach performs four steps.
At first, it computes the hunks between the source and target commits based on an existing ``diff'' tool (line~\ref{overall_algorithm:diffs_start} in Algorithm~\ref{alg:overall_algorithm}).
For each hunk, the approach then determines the positional relationship between the source region and the hunk (lines~\ref{overall_algorithm:hunk_iteration_start} to~\ref{overall_algorithm:hunk_iteration_end}).
The third step computes candidate regions based on the positional relationship (lines~\ref{overall_algorithm:records_iteration_start} to~\ref{overall_algorithm:records_iteration_end}).
Finally, the approach refines the candidate regions to obtain more precise results.
We present each of these four steps in more detail in the following. 

\paragraph{Step 1: Extracting Hunks}
Because different diff algorithms may produce different results, \name{} uses four different algorithms to increase the likelihood of finding the correct candidate region.
The four algorithms are all implemented in the ``git diff'' tool: 
(i) \emph{myers}, the default algorithm, 
(ii) \emph{minimal}, spends extra time to make sure the smallest possible diff is produced, 
(iii) \emph{patience}, tends to be more human-readable, and
(iv) \emph{histogram}, extends the patience algorithm to support low-occurrence common elements.
In addition, we configure ``git diff'' to apply these algorithms at the line-level and at the word-level, resulting in a total of eight diff reports.
\name{} deduplicates these reports and proceeds to extract candidate regions from the deduplicated reports.

\paragraph{Step 2: Determining Positional Relationships}
Given the hunks in a diff report, \name{} determines which hunks are relevant for the source region by analyzing their positional relationship.
We consider a hunk to be \emph{relevant} if it either modifies the source region or affects its location. 
To collect relevant hunks, \name{} considers three major categories of positional relationships between the source region and hunks, as shown in Figure 3, which we further describe in the following:
\begin{enumerate}[label=\roman*]
  \item \emph{Fully covered}: The source region is fully covered by a hunk, i.e., the hunk is relevant. 
  \item \emph{Overlapping}: The source region overlaps with one or more hunks at the top, middle, or bottom, i.e., the hunk is relevant. 
  Specifically, we consider four cases: top overlapping, bottom overlapping, top-bottom overlapping, and middle overlapping.
  \item \emph{Disjoint}: The source region does not overlap with any hunk, i.e., its content remains unchanged. However, if the location of the source region change because of the hunks preceding it, then the hunk is relevant.
\end{enumerate}

\begin{figure}[t]
  \centering
  \includegraphics[width=0.49\textwidth]{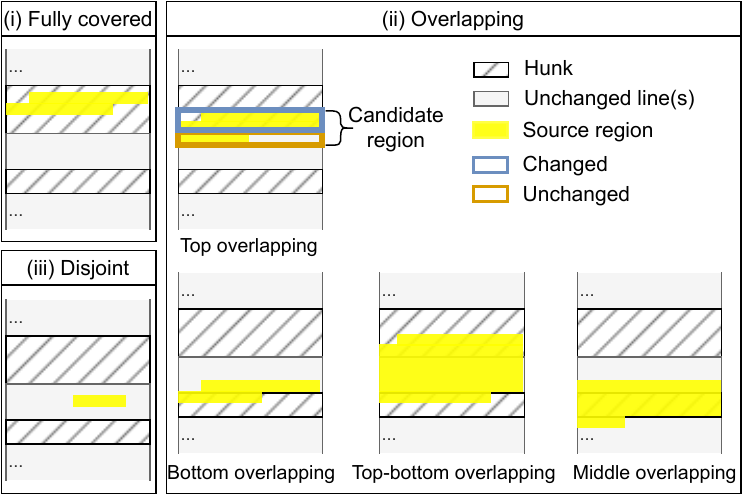}
  \vspace{-0.25in}
  \caption{Positional relationships between a source region and hunks.}
  \label{fig:overlapping_locations}
  \vspace{-0.08in}
\end{figure}

\paragraph{Step 3: Extracting Candidate Regions.}
After identifying the relevant hunks, \name{} extracts candidate regions from them (function $\mathit{getCandidates}$ in line~\ref{overall_algorithm:records_iteration_end} of Algorithm~\ref{alg:overall_algorithm}).
Recall from Definitions~\ref{def:character_range} and~\ref{def:region} that the range of a candidate region is represented as $\mathit{R}_\mathit{target}$=($\mathit{l}_{\mathit{1}}$, $\mathit{c}_{\mathit{1}}$, $\mathit{l}_{\mathit{2}}$, $\mathit{c}_{\mathit{2}}$). 
For hunks that fully cover the source region, \name{} first checks whether the region has been removed in the target commit.
To this end, the approach checks whether the target block ($\mathit{l}_{\mathit{target}}^{\mathit{start}}$, $\mathit{l}_{\mathit{target}}^{\mathit{end}}$) of a hunk is empty, and if so, generates the special candidate region $(\perp, \perp, \emptyset)$.
Otherwise, \name{} continues by determining a coarse-grained candidate range by assigning $\mathit{c}_\mathit{1}$ to a default value 1, $\mathit{c}_\mathit{2}$ to the length of line $\mathit{l}_{\mathit{target}}^{\mathit{end}}$ in the target file, and updating the candidate range as ($\mathit{l}_{\mathit{target}}^{\mathit{start}}$, 1, $\mathit{l}_{\mathit{target}}^{\mathit{end}}$, $\mathit{c}_{\mathit{2}}$). 
Then, the refinement step described in the following paragraph will refine the range to make it more precise, e.g., by excluding unrelated characters. 
For overlapping hunks, e.g., in the case of top overlapping (refer to Figure~\ref{fig:overlapping_locations}), the candidate region is a combination of two parts: one where the source region overlaps with the top overlapping hunk (marked with a blue box), and another that remains unchanged (marked with an orange box). 
\name{} gets a coarse-grained start position of $\mathit{R}_\mathit{target}$ by setting ($\mathit{l}_{\mathit{1}}$, $\mathit{c}_{\mathit{1}}$) as ($\mathit{l}_{\mathit{target}}^{\mathit{start}}$, 1),  
and searches the unchanged part to get the ($\mathit{l}_{\mathit{2}}$, $\mathit{c}_{\mathit{2}}$). Then, the approach runs the refinement step to get a refined start position. 
The other overlapping cases are handled similarly. 
For disjoint hunks, \name{} first initializes candidate ranges to be the same as the source range, and then updates the line numbers by accounting for the changes caused by the hunks located above the source region (line~\ref{overall_algorithm:not_related_update_linenum}). 

\paragraph{Step 4: Refinement of Candidate Regions}
\label{par:refinement}

\begin{figure}[t]
  \vspace{-0.06in}
  \includegraphics[width=0.45\textwidth]{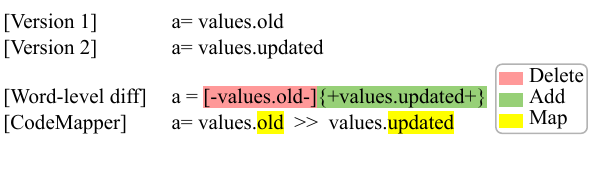}
  \vspace{-0.3in}
  \caption{Refinement of candidate regions.}
  \label{fig:refinement_example}
\end{figure}

To motivate the fourth and final step of our diff-based candidate extraction, 
consider the example in Figure~\ref{fig:refinement_example}.
The code change in the first two lines modifies \code{old} into \code{updated}.
Suppose the source region is \code{old}, i.e., ideally, we want to find \code{updated} as the target region.
However, word-level diff gives the candidate region \code{values.updated}, even though it identifies changes at the word level, the candidate is still insufficiently precise.

To obtain more precise candidate regions, \name{} refines the candidate regions extracted from hunks that fully cover a source region and from overlapping hunks (excluding middle overlapping). 
Intuitively, the refinement step aims to adjust the start and end positions of the candidate regions to more closely match the source region.
Algorithm~\ref{alg:fine_grained_algorithm} summarizes the refinement step, which takes three inputs:
a source range $\mathit{R}_{\mathit{source}}$, a reference hunk $\mathit{H}_{\mathit{ref}}$, and a candidate range $\mathit{R}_{\mathit{target}}$.
The $\mathit{H}_{\mathit{ref}}$ depends on the scenario: 
\begin{itemize}
  \item If a hunk fully covers the source region, then $\mathit{H}_{\mathit{ref}}$ is $\mathit{H}_{\mathit{fully\_cover}}$, which is used to refine both the start and end positions of the candidate region.
  \item For top overlapping and bottom overlapping hunks, the reference hunk is $\mathit{H}_{\mathit{top}}$ and $\mathit{H}_{\mathit{bottom}}$, respectively.
  \item For a top-bottom overlapping hunk, \name{} runs the algorithm twice: $\mathit{H}_{\mathit{top}}$ as $\mathit{H}_{\mathit{ref}}$ for refining the start position and $\mathit{H}_{\mathit{bottom}}$ as $\mathit{H}_{\mathit{ref}}$ for refining the end position.
\end{itemize}

The algorithm shows the details on how to get the refined start position of a candidate region.
To refine the end of the region, \name{} follows a similar approach, but analyzing the $\mathit{H}_{\mathit{ref}}$ in reverse. 
The algorithm starts by identifying in the reference hunk the first line containing a delete operation.
This ensures that the refinement starts from the most significant point of change, aligning with the start line of the source region.
Subsequently, the algorithm uses a word-level diff report to extract text fragments, which represent  individual changes, such as deletions, unchanged segments, and additions in the line.
For each such fragment, \name{} calculates fragment lengths and updates character indices (line~\ref{fine_grained_algorithm:get_precise_char_idx}). 
If the current character index exceeds the starting character position, the algorithm further refines the position at the character-level (line~\ref{fine_grained_algorithm:get_char_idx}). 
At the character-level, the algorithm identifies the overlap between the current fragment and the start of the source region, and then uses this information to exclude any preceding characters not present in the source region. 
Finally, the algorithm outputs a refined candidate range $\mathit{R}'_{\mathit{target}}$. 

For example, in Figure~\ref{fig:refinement_example}, 
\name{} checks the overlap between the fragment \code{values.old} and source characters \code{old}, excludes the \code{values.}, and gives a candidate \code{updated}. 


\SetKwInput{Input}{Input}
\SetKwInput{Output}{Output}
\begin{algorithm}[t]
  \caption{Refinement of candidate ranges.}
  \label{alg:fine_grained_algorithm}
  \small\Input{
  Source region with range $\mathit{R}_{\mathit{source}}\kern-0.2em : (\mathit{l}_{\mathit{s1}}, \mathit{c}_{\mathit{s1}}, \mathit{l}_{\mathit{s2}}, \mathit{c}_{\mathit{s2}})$, reference hunk: $\mathit{H}_{\mathit{ref}}$, 
  candidate range $\mathit{R}_{\mathit{target}}: (\mathit{l}_{\mathit{t1}}, \mathit{c}_{\mathit{t1}}, \mathit{l}_{\mathit{t2}}, \mathit{c}_{\mathit{t2}})$}
  \small\Output{Refined candidate range: $\mathit{R}'_{\mathit{target}}: (\mathit{l}'_{\mathit{t1}}, \mathit{c}'_{\mathit{t1}}, \mathit{l}_{\mathit{t2}}, \mathit{c}_{\mathit{t2}})$}
  
  %
  $\mathit{l}'_{\mathit{t1}} \gets \mathit{getFirstModifiedLine}(\mathit{H}_{\mathit{ref}})$ \label{fine_grained_algorithm:line_init}\\
  \tcp{\small\textbf{Identify \kern-0.2em text \kern-0.2em fragments \kern-0.2em in \kern-0.2em the \kern-0.2em modified \kern-0.2em line}}
  $\mathit{currentSrcCharIdx}, \mathit{cndCharIdx} \gets 0$ \\
  \For{$\mathit{fragment}$ in $\mathit{l}'_{\mathit{t1}}$}{ \label{fine_grained_algorithm:fragment_iteration_start}
    \If{$\mathit{currentSrcCharIdx} \geq \mathit{c}_{\mathit{s1}}$}{ \label{fine_grained_algorithm:char_idx_check_start}
      \tcp{\small\textbf{Further \kern-0.2em refine \kern-0.2em character \kern-0.2em indices}}
      $\mathit{c}'_{\mathit{t1}} \gets \mathit{computeStartChar}(\mathit{R_{source}, fragment}, \mathit{cndCharIdx})$ \label{fine_grained_algorithm:get_char_idx} \\
      \textbf{break}
    }\label{fine_grained_algorithm:char_idx_check_end}
    $\mathit{currentSrcCharIdx}, \mathit{cndCharIdx} \gets \mathit{updateCharIndices(fragment, currentSrcCharIdx, cndCharIdx)}$ \label{fine_grained_algorithm:get_precise_char_idx}
  } 
  $\mathit{R}'_{\mathit{target}} \gets (\mathit{l}'_{\mathit{t1}}, \mathit{c}'_{\mathit{t1}}, \mathit{l}_{\mathit{t2}}, \mathit{c}_{\mathit{t2}})$ \label{fine_grained_algorithm:get_fine_grained_range}\\
  \Return{$\mathit{R}'_{\mathit{target}}$}
\end{algorithm}

\subsubsection{Movement Detection}
\label{section:movement_detection}

Developers sometimes move code fragments via cut-and-paste from one location to another.
Unfortunately, git diff may not always detect such movements, which could lead \name{} to miss the correct target region.
For example, the \code{print(s)} in line 9 is moved to line~14 in Figure~\ref{subfig:motivation_track_regions}, but Figure~\ref{subfig:motivation_diff} misidentifies it is deleted.
To address this issue, \name{} includes a movement detection technique that identifies moved code fragments, which is loosely inspired by prior work on generating edit scripts~\cite{higo2017generating}.
We consider two kinds of movements:
(i) \emph{Vertical movements:} Lines are cut-and-pasted to another location, without modifying the lines. 
(ii) \emph{Horizontal movements:} Lines are moved inside or outside a structural unit, such as a block of code.

To identify otherwise missed movements, \name{} marks a source region as potentially moved if it is fully deleted according to line-level git diff. 
For any such potentially moved source regions, the approach checks all hunks to find a corresponding movement. 
Specifically, it checks for hunks that contain all lines of the source region as newly added lines, which detects vertical movements, and for hunks that contain the source region's lines except for whitespace, which detects horizontal movements. 
If such a match is found, it adds those ``added'' lines as a candidate region. 
This step is summarized as $\mathit{detectMovements}$ in line~\ref{overall_algorithm:fully_delete_checking_end} of Algorithm~\ref{alg:overall_algorithm}.

\subsubsection{Text Search}
\label{section:search}
As a third technique to identify candidate target regions, \name{} searches for the exact string from the source region in the target file. 
Text searching is typically useful for relatively small source regions, e.g., a single variable or a short expression, which may get lost in a larger hunk.
The approach adds any exactly identical occurrences of the source region in the target file as candidate regions  (Algorithm~\ref{alg:overall_algorithm}, line~\ref{overall_algorithm:search_candidates}).

\subsection{Target Region Selection}
\label{section:target_selection}

The approach described so far (phase 1) results in three sets of candidates: diff-based, movement-detected, and searched. 
Phase 2 selects the most likely target region from these candidates.
At first, \name{} merges and deduplicates these sets.
Next, it computes the similarity (details in next paragraph) between the source region and each candidate region.
Finally, the approach selects the candidate with the highest similarity score as the target region.
If multiple candidates have the same priority, \name{} heuristically prioritizes the diff-based candidates over the movement-detected candidates, and then the searched candidates.
A benefit of ranking all candidates and then selecting the most similar one is that we avoid the need to set a similarity threshold~\cite{Mossad_threshold2020,buggy_patch_similarity2021}, which can be challenging to determine and may lead to incorrect results~\cite{similarity-review2019}.


To compute the similarity of a source region and a candidate region, \name{} considers both the content of the regions and their context.
As a similarity metric, we use Levenshtein similarity, which is based on the number of character-level edits required to transform one string into another.
Other metrics, e.g., based on neural code embeddings could be easily integrated into \name{}, but we have chosen Levenshtein similarity for its simplicity.

Because the correct target region may modify the code in the source region, computing the similarity only between these two regions may give misleadingly low similarities.
Instead, \name{} also considers the context of these regions, i.e., a fixed number of unchanged lines before and after these regions.
Function $\mathit{computeSimilarity}$ (Algorithm~\ref{alg:overall_algorithm}, line~\ref{overall_algorithm:similarity_computation}) receives a source region $\mathit{G_{source}}$ and a set of candidates as inputs.
It iterates through each candidate region $\mathit{G_{target}}$ and computes its similarity to the source region as follows: 
$$\mathit{simil}(\mathit{addContext}(G_{source}), \mathit{addContext}(G_{target}))$$
where the function $\mathit{addContext()}$ retrieves the context for each region.
One exception to the above is that \name{} does not add any context when computing the similarity for movement-detected candidate regions.
The rationale is that moved code fragments are likely to have different context lines in the source and target files, as the surrounding lines are often changed when the region is relocated.
We set the context size as 15 lines before and 15 lines after the region.
This size strikes a balance: A smaller size risks incorrect candidate regions occasionally sharing the same context lines, while a larger size may reduce the impact of the actual region.
Section~\ref{sec:context_sizes} evaluates the impact of the context size on our results.

\section{Evaluation}

\subsection{Research Questions}
\label{section:RQ}
Our evaluation investigates the following research questions:
\begin{enumerate}[label=RQ\arabic*]
  \item Effectiveness: How effective is \name{} in finding the target region for a given source region?
  \item Impact of parameters and components: How do different parameters and components impact \name{}'s effectiveness?
  \item Efficiency: How much execution time does \name{} take?
\end{enumerate}

\subsection{Experimental Setup}
\subsubsection{Datasets}
\label{section:datasets}

We apply \name{} to three datasets, which are all based on real-world commits.
Table~\ref{t:datasets} provides an overview of the datasets, which we explain in more detail in the following.



\input{tables/dataset_replacement.tex}

\paragraph{Annotated Data~A and B}
As there is no existing dataset for the code mapping problem, we create and manually annotate two new datasets, called Data~A and Data~B.
Each dataset consists of 100 pairs of a source region and its corresponding target region.
We start by selecting 20 projects on GitHub based on programming language and popularity.
Specifically, we pick ten programming languages that are popular on GitHub, randomly select two popular projects for each language, and then randomly sample from the latest 200 commits of each project.
To avoid trivial code mapping tasks, we select only files that are modified in the chosen commits. 
To select source and target commits, we define \emph{commit distance} as the number of commits between source commit and target commit that change the specified file.
We sample half of the source-target pairs from neighboring commits, i.e., a commit distance of one, and the other half from non-neighboring commits with a maximum distance of five.
Additionally, we randomly choose for each pair a mapping direction (forward or backward), i.e., whether the old or the new code version serves as the source. 

Next, we annotate source-target pairs, covering four change operations: no change (25\%) and change, move, and delete (75\%). 
For Data~A, we first manually select a source region and then its corresponding target region.
To encompass various scenarios, we annotate source regions of different sizes (each contributing 20\%): 
single identifier or word, single expression or part of a sentence, single line, multiple lines, and structural unit. 
For Data~B, to further reduce bias, we select source regions automatically and annotate only the corresponding target regions manually.
To prevent meaningless source ranges, e.g., a generated source region starting in the middle of a token, we compute the Abstract Syntax Tree (AST) of each selected file with tree-sitter, and then generate source regions with three strategies (each contributing 33.3\%):
(i) randomly select a single node as a source region,
(ii) randomly select a sequence of consecutive sibling nodes as a source region, and
(iii) randomly select a sequence of consecutive lines as a source region.
The generated source regions vary in size, ranging from individual tokens, over partial and  full lines, to multiple consecutive lines.
Given the automatically generated source regions, we manually annotate the target regions. 
For both Data~A and Data~B, two of the authors annotate each example individually.
One annotator is a PhD-level researcher, the other is a senior researcher.
Both have over five years of experience in software development and extensive experience in software evolution.
The annotation process leads to an initial agreement on 94/100 and 92/100 examples, followed by a discussion during which the annotators resolve any disagreements.

\paragraph{Data from Suppression Study}
\name{} is valuable not only for helping developers find target regions based on a given source region, but also for enabling empirical studies on software evolution. 
For example, our approach could be used to support empirical studies on the evolution of specific kinds of code elements, such as type annotations~\cite{fse2022_type_study}, comments~\cite{fse25_suppression_study}, or suppressions of static analysis warnings~\cite{fse25_suppression_study}.
To evaluate this usage scenario, we reuse a dataset from a study on suppressions in Python code bases~\cite{fse25_suppression_study}. 
The term ``suppression'' here refers to the intentional practice of ignoring certain warnings generated by static analysis tools by adding an annotation or special comment into the code.
The existing dataset contains histories of suppressions, i.e., a sequence of code changes that tracks a specific suppression across the lifetime of a project.
The original dataset has been created by a custom tool that tracks suppressions in Python code bases, and has been validated for correctness by the original authors~\cite{fse25_suppression_study}.
We here evaluate whether \name{} could have been used to map suppressions in Python code bases from one commit to another, which would have saved the effort of creating a custom tracking tool.

The suppression study data consists of 187 source-target pairs, each representing a change that modifies a suppression in a Python file.
We target all code changes used in the original study where (i) the file containing the suppression still exists in the new version and (ii) the line containing the suppression is involved in a change.
In addition, the original dataset contains changes where entire files are deleted and where the line containing the suppression is not changed at all.
We exclude these changes here because those suppressions are trivial to map.

\paragraph{Data from Prior Work on Code Tracking}

To further evaluate generalizability, we build on a dataset from CodeTracker, i.e., prior work on code tracking~\cite{10.1145/3540250.3549079,Hasan:TSE:2024:CodeTracker2.0}. 
Since our code mapping task differs from their code tracking task (cf.\ end of Section~\ref{sec:terminology}), we process their data to retrofit it to the code mapping task. 
The original dataset contains ground truth histories of variables, blocks, and methods in Java projects, where each history consists of a sequence of commits $[c_1, c_2, ..., c_n]$ that modify the code element.
To transform this data into a code mapping dataset, we extract all pairs of commits that are consecutive in the given code element history, i.e., $(c_i, c_{i+1})$ for $i = 1, ..., n-1$.
Note that the project's history may contain many other commits in between such a pair of commits, which affect the code around the code element in question, but not the code element itself.
Furthermore, to fit the existing data into our code mapping task, we require the code region (Definition~\ref{def:character_range}) of the code element to be specified in both commits.
To this end, we parse the corresponding code files and extract the detailed location of the changed code elements.
Following the original work~\cite{10.1145/3540250.3549079,Hasan:TSE:2024:CodeTracker2.0}, the task is to map the code regions backward in time, and we focus on the ``test'' partition of the entire dataset.
After this processing, we obtain a total of 2,005 source-target pairs, including 530 for variables, 949 for blocks, and 526 for methods. 
\subsubsection{Baselines}

We compare \name{} against two baselines: line-level git diff and word-level git diff, which we refer to as diff\textsubscript{line} and diff\textsubscript{word}, respectively.
We select these baselines because they are widely used in practice and because our approach builds on them.
%
Given a source region, we use git diff to identify its relevant hunk(s). Then, we use the corresponding hunk(s) in the target commit as the target region. This process mimics what a developer trying to map a code region would do using git diff. 

In addition to reusing data from prior work on code tracking (Section~\ref{section:datasets}), we also considered directly comparing against work on code tracking.
The two most recent approaches are CodeTracker~\cite{10.1145/3540250.3549079,Hasan:TSE:2024:CodeTracker2.0} and FinerGit~\cite{2020FinerGit}.
However, these approaches are not directly comparable to \name{} because they address a different problem, namely the problem of finding the commits in which a method, variable, or code blocks gets changed (cf.\ Section~3.2 of \cite{10.1145/3540250.3549079} and Section~5.1 of \cite{2020FinerGit}).\footnote{CodeTracker is also evaluated on its ability to determine the kind of code change, e.g., whether method's documentation or parameters have changed, but this is not the focus of our evaluation.}
In contrast, our work assumes a source commit and a target commit to be given, and aims to identify the target region in the target commit that corresponds to a source region in the source commit.


\subsubsection{Evaluation Metrics}

We evaluate an identified target region by comparing it with a ground truth using several metrics. 
First, we check whether the target region \emph{overlaps} with the ground truth.
If yes, we further distinguish between an \emph{exact match}, i.e., the predicted target region is exactly the same as the ground truth and a \emph{partial overlap}. 
In addition, we compute for all target regions the metrics in Figure~\ref{fig:metrics}.
Finally, for partial overlaps, we compute the \emph{character distance} between the target region and the ground truth.
To compute this, we view the target region and the ground truth as substrings in the sequence of characters of a file, ranging from indices $i$ to $j$ and $i'$ to $j'$, respectively.
The character distance is defined as shown in Figure~\ref{fig:metrics}. 
For example, if one region starts at 20 and ends at 55, and the other region starts at 18 and ends at 63, then the character distance is $|20-18| + |55-63| = 10$.

\subsection{RQ1: Effectiveness}

\input{tables/ressult_rq1_1.tex}
\input{tables/ressult_rq1_2.tex}

\subsubsection{Results on Data~A}

The leftmost block (Data~A block) of Table~\ref{tab:result_anno_supp} shows the results of mapping code regions in the Data~A.
As there are 100 tasks in the dataset, all percentages mentioned in the following equal the absolute numbers.
\name{} identifies 95\% overlapping target regions, compared to diff\textsubscript{line} with 89\% and diff\textsubscript{word} with 86\%. 
Specially, our approach successfully identifies 77\% exactly matched target regions, outperforming both diff\textsubscript{line} (43\%) and diff\textsubscript{word} (54\%). 
The exact matches include eight cases where both baselines fail to identify movements, while \name{} provides exact matches by utilizing its movement detection.
The character distance indicates that the 18 partial overlaps of \name{} are 55.2 characters away from the expected regions, on average, which is clearly better than both baselines (147.0 and 71.1 characters).
That is, even when the approach cannot exactly identify the target region, it still provides more precise regions than the baselines.

\begin{figure}[t]
  \centering
  \includegraphics[width=0.49\textwidth]{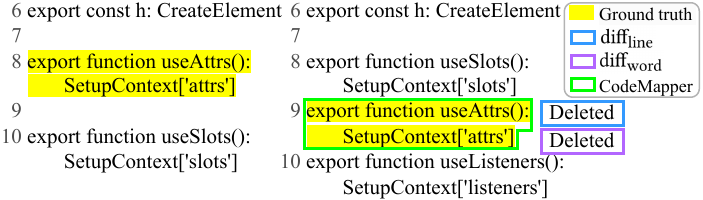}
  \vspace{-0.28in}
  \caption{A source region involved in a vertical movement. Only \name{} finds an exact match of the target region.} 
  \vspace{-0.01in}
  \label{fig:suppression_involves_in_movement} 
\end{figure}

\begin{figure}[t]
  \centering
  \includegraphics[width=0.49\textwidth]{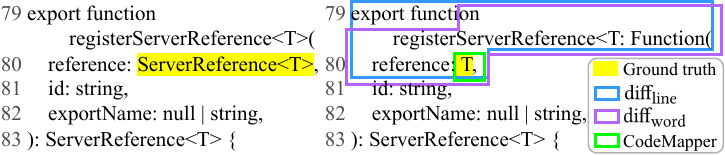}
  \vspace{-0.28in}
  \caption{A close match found by refining candidate regions.} 
  \vspace{-0.02in}
  \label{fig:minimize_character_distance} 
\end{figure}

Figure~\ref{fig:suppression_involves_in_movement} shows an example where a source region is moved vertically, with lines 8 and 10 swapped. 
The source and expected target regions are highlighted in yellow.
Both baselines misidentify the source region as deleted. 
Instead, \name{} benefits from its movement detection (Section~\ref{section:movement_detection}) and finds the exact target region, highlighted in green. 

Figure~\ref{fig:minimize_character_distance} illustrates the importance of refining candidate regions. 
Both baselines report a region much larger than the expected target region.
Although \name{} does not achieve an exact match, its output is very close to the ground truth.
This is mainly because it successfully refines a diff-based candidate (Step 4 of Section~\ref{par:refinement}). 

\subsubsection{Results on Data~B}  

The Data~B block of Table~\ref{tab:result_anno_supp} presents the results on this dataset. 
\name{} identifies 71\% exact matches, surpassing diff\textsubscript{line} at 44\% and diff\textsubscript{word} at 54\%. 
It also achieves the lowest average character distance of 18.5 characters. 
diff\textsubscript{word} has a larger average character distance than diff\textsubscript{line} because character distance is calculated only for partial overlaps, and diff\textsubscript{line} has fewer partial overlaps. 
This indicates that diff\textsubscript{line} incorrectly maps some cases, while diff\textsubscript{word} provides a partial overlap. 
Additionally, \name{} achieves the highest precision (0.883) and the highest F1-score (0.882). 
However, diff\textsubscript{line} attains the highest recall (0.906), followed by \name{} with a recall of 0.892. 
This difference occurs because diff\textsubscript{line} always assigns a line-level target region, even when only a single token within the line is affected. 
These results, consistent with those for Data~A, show that \name{} is effective at finding target regions for varying sizes of source regions across different programming languages.

\begin{figure}[t]
  \centering
  \includegraphics[width=0.49\textwidth]{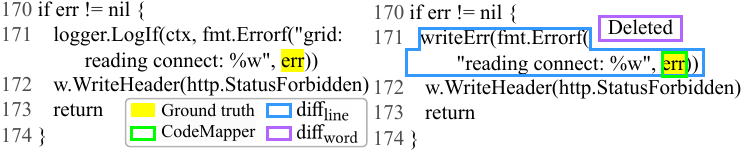}
  \vspace{-0.28in}
  \caption{An exact match found by searching texts.} 
  \label{fig:search_exact_match} 
\end{figure}

Figure~\ref{fig:search_exact_match} provides an example where the line containing the source region has changed, but the source region itself remains the same. 
diff\textsubscript{line} reports the entire modified line (highlighted in blue) as the target region, 
and diff\textsubscript{word} identifies the change as a deletion. 
Meanwhile, \name{} benefits from text search (Section~\ref{section:search}) and accurately detects an exact match, marked with green color.

\subsubsection{Results on Suppression Study Data} 
\label{par:rq1_suppression}

The ``Suppression study data'' block of Table~\ref{tab:result_anno_supp} summarizes the results of mapping Python suppressions.
\name{} identifies 151 (80.7\%) exact matches, outperforming both diff\textsubscript{line} and diff\textsubscript{word}, which both identify only 41 (21.9\%) exact matches.
This improvement of 58.8 absolute percent points is mainly due to refinements of candidate regions, the movement detection, and the text search. 
\name{} also significantly reduces the character distance to 29.2, showing an improvement over diff\textsubscript{line} (199.4) and diff\textsubscript{word} (194.1). 
It also achieves the highest recall at 0.834, the highest precision at 0.824, and the highest F1-score at 0.827. 
Excluding the 21.9\% exact matches found by all three approaches, the difference between the partial overlaps arises because diff\textsubscript{word} focuses on detailed intra-line changes, whereas diff\textsubscript{line} considers entire lines. 
Entire lines have a higher chance of overlapping with the expected target regions. 
In contrast, by trying to identify detailed changes, diff\textsubscript{word} achieves a lower character distance and a higher precision of 0.377 (diff\textsubscript{line} with 0.351).
  
\begin{figure}[t]
  \centering
  \includegraphics[width=0.49\textwidth]{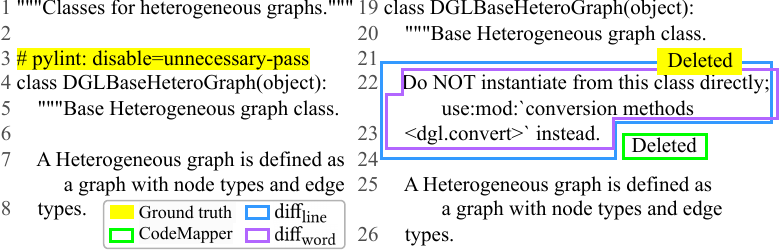}
  \vspace{-0.26in}
  \caption{Only \name{} correctly recognizes that the source region gets deleted.} 
  \label{fig:only_at}
\end{figure}

Figure~\ref{fig:only_at} shows an example where a suppression is deleted in the target file.
The baselines give incorrect mappings for the source region. 
This could mislead an empirical study on suppression evolution by suggesting that the suppression is moved or changed. 
Instead, \name{} correctly identifies the deletion, matching the expected ground truth.

\subsubsection{Results on CodeTracker data}
Table~\ref{tab:result_codetracker_data} presents the mapping results on the CodeTracker data.
\name{} identifies the most exact matches and the smallest character distances for partial overlaps across all three types of program elements. 
Specifically, \name{} achieves exact match rates of 75.7\% for variables, 81.8\% for blocks, and 94.5\% for methods, outperforming the baselines by up to 33.2 percentage points.
Compared to the other three datasets, the results show relatively larger character distances, which can be attributed to the characteristics of CodeTracker data:
Variables are generally small and may occur multiple times, causing partial overlaps to cover larger code regions. 
For blocks and methods, since their source and target regions are much larger (as shown in Table~\ref{t:datasets}), a partial overlap that fails to correctly identify the expected start or end lines may include multiple extra lines, which increases the character distance.
In line with the other datasets, \name{} achieves the highest precision and F1-score, showing its contribution over the state of the art.

\subsubsection{Analysis of Incorrect Target Regions} 
\label{par:incorrect_analysis} 
To gain insights into cases where \name{} selects incorrect target regions, we analyze 219 corresponding cases from the four datasets. 
We observe three recurring reasons for incorrect target regions: 
(i) \emph{Semantic changes not accurately captured by diff}:
This category includes 98 cases where semantic changes are not accurately captured by diff. 
For example, in one case, \code{x:CompileBindings="True"} was removed, and \code{ClassModifier="internal"} was added, which diff incorrectly reported as a match, even though the source region was actually deleted.
(ii) \emph{Coincidental occurrences of the source region}:
This comprises 75 cases where coincidental occurrences of the source region appear at unrelated locations. 
\name{} mitigates this kind of mis-selection by using unchanged lines as context. 
(iii) \emph{Detecting file renames and movements}:
Detecting file renames and movements is generally challenging, especially when a file undergoes significant changes. 
\name{} uses git log to detect file renames and movements, which sometimes (46 cases) fails to identify the correct file in the target commit.
Language-specific techniques for detecting file names and movements, such as RefactoringMiner~\cite{TSE22RefactoringMiner2.0}, as used by CodeTracker~\cite{10.1145/3540250.3549079,Hasan:TSE:2024:CodeTracker2.0}, could mititgate this issue, but is inherently limited to the programming language it supports.



\subsubsection{Impact of Programming Language}

To investigate whether the mapping results differ across programming languages, we analyze the results of Data~A and~B grouped by language. 
For Data~A, F1-scores range from 0.698 to 1.000, with examples including Python (0.950), Java (0.853), and Ruby (1.000). 
For Data~B, F1-scores range from 0.750 to 1.000, with examples including JavaScript (0.870), C++ (0.871), and Go (0.980).
We attribute these variations in F1-scores to the randomized procedure of creating these datasets.
For instance, a particularly complex source region appearing in one language could lower the F1-score for that language.
Additionally, the fixed annotation ratios (e.g., 25\% no-change, 50\% forward time-order) and the inclusion of non-code files (e.g., Markdown and YAML) contribute to an unbalanced number of source regions across languages.
Although the possibility of language-specific trends cannot be entirely excluded, the present results do not provide consistent or statistically significant evidence to support this conclusion.

\begin{finding}
  \name{} is effective at mapping code regions for various kinds of source regions across different programming languages.  
  The approach identifies 77.0\% exact matches for Data~A, 71.0\% for Data~B, 80.7\% for the suppression study data, 75.7\% for CodeTracker data variable, 81.8\% for block, and 94.5\% for method, surpassing the best available baselines by 1.5--58.8 absolute percent points.
\end{finding}

\subsection{RQ2: Impact of Parameters and Components of \name{}}
\label{sec:answer_to_RQ2}
To better understand the impact of different parameters and components of \name{} on its effectiveness, we conduct two ablation studies: one examining different context sizes and another analyzing specific components of the approach.

\subsubsection{Context Size}
\label{sec:context_sizes}
\begin{figure}[t]
  \centering
  \includegraphics[width=0.44\textwidth]{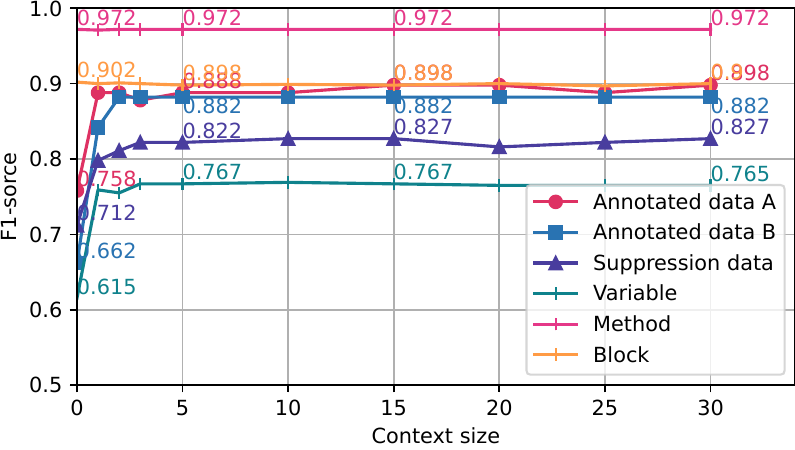}
  \vspace{-0.14in}
  \caption{F1-scores with different context sizes.}
  \label{fig:f1s}
\end{figure}

\name{} uses code context to help identify target regions (Section~\ref{section:target_selection}). 
We evaluate different context sizes and present the corresponding F1-scores in Figure~\ref{fig:f1s}. 
We plot the F1-score as it represents the overall trend and in line with the other metrics.
The results show that using a context size of five or more provides better results than small values, meaning that \name{} effectively leverages contextual information.
At the same time, context sizes between 5 and 20 provide similar effectiveness, i.e., the approach is robust to minor changes of this parameter.
We use 15 as the default context size in all other experiments.

\subsubsection{Importance of Different Components}

\begin{figure*}[t]
  \centering
  \includegraphics[width=\linewidth]{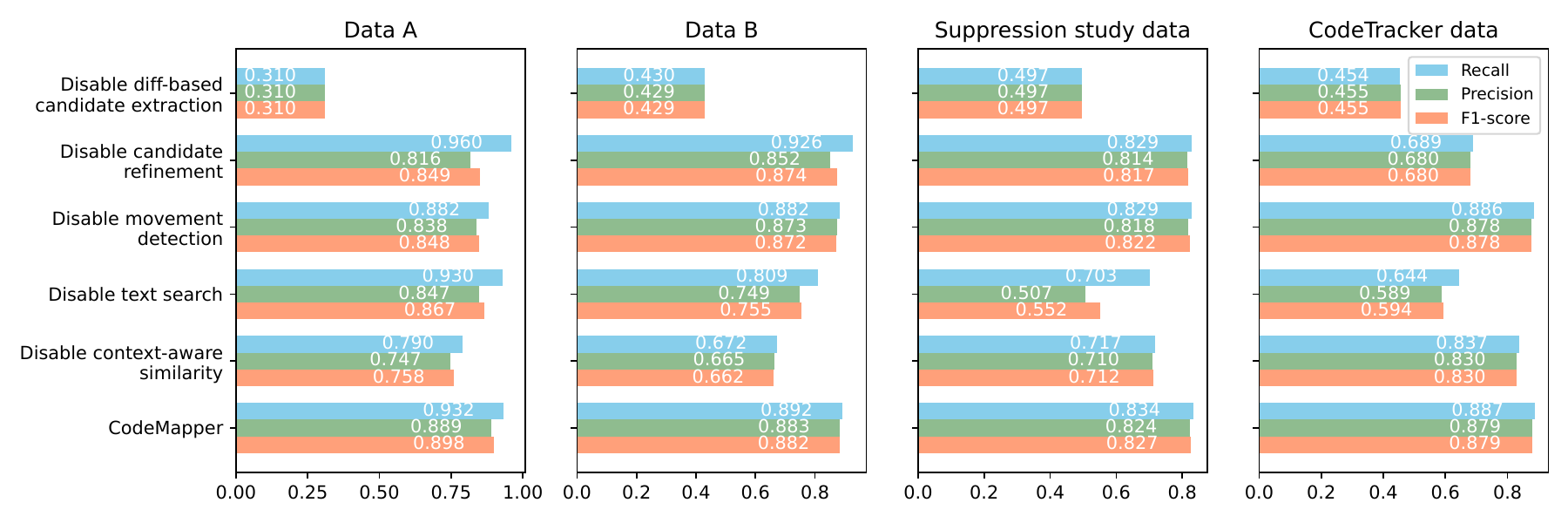}
  \vspace{-0.32in} 
  \caption{Ablation study of different components.}
  \label{fig:ablation_component}
\end{figure*}

We study five variants of the approach, each with one part disabled.
Specifically, we disable diff-based candidate extraction, refinement of candidate regions, movement detection, text search, and the context-aware similarity metric. 
Figure~\ref{fig:ablation_component} shows the results of the ablation study for the four datasets.
The results show that diff-based candidate extraction is the most critical component, as disabling it leads to the largest decrease in all metrics. 
This is unsurprised since \name{} relies on diff reports for basic line mappings.
Without it, the approach relies solely on movement detection, which helps only when the code region actually was moved, and text search, which helps only when the code region is unchanged.
Disabling text search leads to a relatively large drop for the suppression study data and the CodeTracker data, with F1-scores decreasing from 0.827 to 0.552 and from 0.879 to 0.594, respectively. 
This is mainly because text search remains effective at mapping unchanged code regions, such as some suppressions and variables, highlighting its contribution to the overall performance. 
Disabling the other components also reduces the effectiveness, either on some or all datasets.
For example, disabling movement detection significantly reduces recall on Data~A, missing seven moved code regions that \name{} identifies as an exact match.
Another example is that refining candidate regions increases precision, e.g., from 0.816 to 0.889 on Data~A, which is exactly what this component is designed for.

We further analyze the contribution of the four git diff algorithms. 
The results indicate that \name{} identifies 2,405 candidates solely due to one of the three non-default algorithms: 
1,955 from \emph{patience}, 305 from \emph{histogram}, and 145 from \emph{minimal} (Section~\ref{section:diff_based_candidate_extraction}). 
From these 2,405 candidates, phase~2 of our approach selects 74 as target regions, leading to 22 exact matches and 23 partial overlaps. 
Given the relatively low computational cost of running multiple diff algorithms (Section~\ref{sec:efficiency}), \name{} uses all four algorithms to increase the chance of collecting more candidate regions.

\begin{finding}
  Considering contextual lines helps in selecting the most likely target region, and \name{} is robust to minor changes of context size. 
  Each component of \name{} enhances its overall effectiveness, with the diff-based candidate extraction and text search being the most critical.
\end{finding}
\vspace{-0.06in}

\subsection{RQ3: Efficiency of \name{}}
\label{sec:efficiency}

\begin{figure}[t]
  \vspace{-0.06in}
  \centering
  \includegraphics[width=0.49\textwidth]{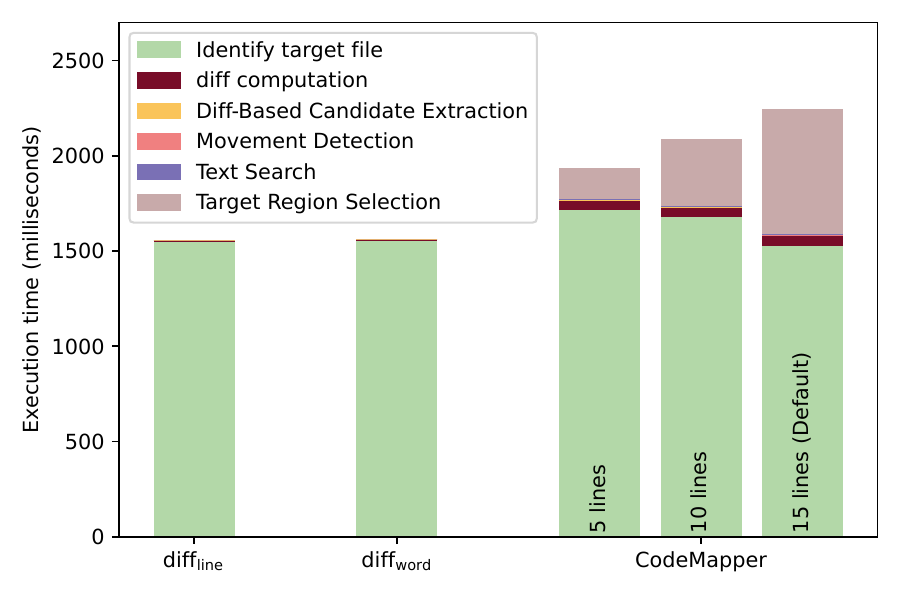}
  \vspace{-0.35in}
  \caption{Execution time across the four datasets.} 
  \vspace{-0.22in}
  \label{fig:execution_time}
\end{figure}

To measure efficiency, we record the execution times of \name{}. 
Figure~\ref{fig:execution_time} visualizes the overall execution time, averaged over five runs. 
In its default configuration, \name{} takes 2,327 milliseconds to compute a target region.
Because the baselines do not first compute candidate regions and then select the best target region, they are faster, with an average of 1,624 and 1,628 milliseconds for diff\textsubscript{line} and diff\textsubscript{word}, respectively.
Despite the increased effort performed by \name{}, our approach is fast enough for interactive usage.
Because smaller context sizes than our default of 15 also provide competitive results (Figure~\ref{fig:f1s}), Figure~\ref{fig:execution_time} also shows the execution time with smaller context sizes.
Reducing the context significantly reduces the execution time, offering users a knob for trading slightly reduced effectiveness for even higher efficiency.

The breakdown of where time is spent, as shown with distinct colors in Figure~\ref{fig:execution_time}, reveals that \name{} takes 1,665 milliseconds to compute candidates for a source region and 661 milliseconds to select a target region.
While computing candidates, the approach runs diff eight times (Section~\ref{section:diff_based_candidate_extraction}), whereas the baselines use only one diff result.
The time taken for ``target region selection'' is mostly due to computing the Levenshtein distances.

\begin{finding}
  \name{} takes an average of 2,327 milliseconds to identify a target region, which, despite being slower than the baselines, is sufficiently fast for interactive usage.
  Reducing the context size further reduces the execution time.
\end{finding}

\section{Threats to Validity}
The primary threat to validity concerns the datasets.
Data~A is created by the authors, which may introduce bias.
To reduce this bias, Data~B starts from randomly selected source regions.
Each of the 200 examples has been annotated independently by two authors, followed by a discussion to reach consensus.
With only 6 and 8 initial disagreements in Data~A and B, respectively, we consider the annotation results highly reliable.
A related threat is using the suppression study data~\cite{fse25_suppression_study} as ground truth.
To mitigate this, we randomly inspect 20 history pairs, all of which we found correct.
Another potential threat is the limited size of the first three datasets.
We mitigate this by also evaluating \name{} on the larger CodeTracker data~\cite{10.1145/3540250.3549079,Hasan:TSE:2024:CodeTracker2.0}, which was created independently.
Finally, our approach uses git diff as a component and baseline, but other diff implementations exist.
We selected git diff because it is widely used and represents the state of the art in diff computation.


\section{Related Work}

\paragraph*{Studies of Developer Needs and Behaviors}
LaToza et al.~\cite{Hard-to-answer2010} conduct a survey to identify questions about code that developers find challenging to answer. 
Their findings indicate that developers are interested in code histories at the code snippet level, whereas many existing tools require navigating through all changes at the file level to locate specific smaller changes of interest.
\name{} addresses this gap by enabling developers to map customized code regions from one commit to another.
Lin et al.~\cite{finegrainedchangetypes2016} investigate the characteristics of fine-grained change types and find that changes to functions and statements are the most prevalent. 
Codoban et al.~\cite{Codoban2015} study why developers check program histories.
They observe different motivations, which can be categorized into three groups: 
finding specific commits, being aware of changes, and finding safe points for backtracking.
Our approach could help developers in all three categories by providing a more detailed view of code changes.

\paragraph*{Tracking of Code Elements}
Previous work~\cite{icse_2021_codeshovel,10.1145/3540250.3549079,Hasan:TSE:2024:CodeTracker2.0} focuses on tracking specific program elements across version histories.
CodeShovel~\cite{icse_2021_codeshovel} supports Java methods.
CodeTracker~\cite{10.1145/3540250.3549079} supports Java methods and variables, while CodeTracker 2.0~\cite{Hasan:TSE:2024:CodeTracker2.0} extends this to blocks.
Our work differs by addressing the code mapping problem, where two commits are given, and the task is to find the target region for a source region.
We also present a language-agnostic approach applicable to arbitrary code regions.
FinerGit~\cite{2020FinerGit} and Historage~\cite{2011Historage} utilize a finer-grained Git repository structure by reorganizing Java methods into individual files for precise tracking.
FinerGit enhances Historage's ability by breaking down method lines into single tokens for token-based diff computation.
However, this reorganization is time-consuming and increases disk space usage in large repositories.
In contrast, \name{} achieves character-level tracking on existing repositories.


\paragraph*{Approaches based on Diffs}
MergeGen~\cite{2023MergeGen} divides line-level conflicts into smaller, more precise conflicts.
Language-independent diff techniques~\cite{2013LHDiff, canfora2008tracking, 2008Trackinglocations} operate on textual code representation.
HyperDiff~\cite{2023HyperDiff} computes scalable AST-level diffs over large code histories.
Unlike our work, these approaches cannot track specific regions of interest.
Matsumoto et al.~\cite{2019BeyondGumTree} combine AST structures and line differences to improve diff comprehension.
Higo et al.~\cite{higo2017generating} consider copy-and-paste actions to enhance diff understandability.
These works emphasize either language independence or fine-grained change capture.
In contrast, \name{} addresses both by employing text-based diffs and character-level regions.

\paragraph*{Reasoning about Code Changes}
Etemadi et al.~\cite{AugmentingDiffs2023} enhance code diffs with runtime information.
DiffSearch~\cite{DiffSearch2023} allows developers to query code changes.
B2SFINDER~\cite{B2SFinder2019} addresses binary-to-source code matching.
Wu et al.~\cite{DIFFBASE2021} manage evolving software artifacts using differential facts for unified analysis.
LibvDiff~\cite{LibvDiff2024} identifies software versions by leveraging symbol information and detecting function-level differences.
These works emphasize the importance of detailed code analysis and suggest techniques for future enhancements of \name{}.
Integrating dynamic information may further improve mapping accuracy in future versions of \name{}.

\paragraph*{Reasoning about Code Line Evolution}
Meta-differencing~\cite{ICSM2004_Meta_differencing} represents source code with abstract syntax information, enabling queries and searches for changes but requires parsing code into an AST and operates only at the line level.
Other line-level approaches identify related changes~\cite{MSR2006_cochange}, detect change types applied to lines over time~\cite{MSR2007_line_levinshtein_dist}, and track related code lines across versions~\cite{FSE12_History_slicing, ICSE17_Fuzzy_history_slicing}.
In contrast, \name{} maps code regions between commits.
Kim et al.~\cite{MSR2006_element_matching_survey} surveys matching techniques for cross-version analysis, noting limitations such as applicability only to parsable projects and fixed granularity.
Our work addresses these by supporting arbitrary code regions and being language-agnostic.

\section{Conclusion}

\name{} empowers developers to map arbitrary code regions between commits. 
Given a source region, it combines novel techniques to identify candidates for the target region and selects the most likely one based on code similarity. 
Like the git diff tool, \name{} works across all text-based programming languages. 
Unlike git diff, our approach allows developers to focus on specific code regions rather than showing all changes at once. 
Our evaluation across hundreds of code changes in ten languages demonstrates the effectiveness and efficiency of \name{}.

\section*{Data~Availability}
\url{https://github.com/sola-st/CodeMapper}

\begin{acks} 
Supported by the European Research Council (ERC; grant
agreements 851895 and 101155832) and by the German Research Foundation
(DFG; projects 492507603, 516334526, and 526259073).
\end{acks}

\bibliographystyle{ACM-Reference-Format}
\bibliography{references, referencesMP}

\end{document}

%% file: listing_pls/listing_python.tex
\definecolor{mono1}{RGB}{56,58,66}
\definecolor{mono2}{RGB}{105,108,119}
\definecolor{mono3}{RGB}{156,158,171}

\definecolor{cyan}{RGB}{1,132,188}
\definecolor{blue}{RGB}{64,120,242}
\definecolor{purple}{RGB}{166,38,164}
\definecolor{green}{rgb}{0.05, 0.5, 0.06}
\definecolor{red1}{rgb}{1.0, 0.89, 0.88} 
\definecolor{red2}{RGB}{202,18,67} 
\definecolor{orange1}{RGB}{152,104,1}
\definecolor{orange2}{RGB}{193,132,1}
\definecolor{hlyellow}{rgb}{1.0, 0.94, 0.0}
\newcommand\digitstyle{\color{orange1}}
\definecolor{grannysmithapple}{rgb}{0.74, 0.89, 0.65}
\definecolor{lightblue}{rgb}{0.68, 0.85, 0.9}
\definecolor{codegreen}{rgb}{0,0.6,0}
\definecolor{codegray}{rgb}{0.5,0.5,0.5}
\definecolor{codepurple}{rgb}{0.58,0,0.82}

\makeatletter
\newcommand{\ProcessDigit}[1]
{%
  \ifnum\lst@mode=\lst@Pmode\relax%
   {\digitstyle #1}%
  \else
    #1%
  \fi
}
\lstdefinelanguage{Python}{
	basicstyle=\ttfamily\footnotesize,
	classoffset=0,
	morekeywords={def, await, import, lambda, break, while, in, return, else, pass, for, del, from, global, not, class, as, elif, if, yield, assert},
	keywordstyle=\color{purple!70!black},
	classoffset=1,
	morekeywords={random},
	keywordstyle=\color{red1},
	classoffset=2,
	morekeywords={True, False, None},
	keywordstyle=\color{orange1},
	classoffset=3,
	morekeywords={print, type, str, float, Exception,dict, int,  raise, list,sum},
	keywordstyle=\color{cyan!80!black},
	classoffset=4,
	morekeywords={round,remove, Custom,MyObj, get_betraying_probability, choice, range, read_csv, Characters, Location_helper},
	keywordstyle=\color{cyan},
	identifierstyle=\color{mono1},
	sensitive=false,
	comment=[l]\#,
	stringstyle=\color{brown}\ttfamily,
	commentstyle=\itshape\color{mono1!80}\ttfamily,
	numberstyle=\color{mono3}\ttfamily,
	escapeinside={/*@}{@*/},
}

\makeatletter
\newcommand\HL{%
   \gdef\lst@alloverstyle##1{%
     \fboxrule=0pt
     \fboxsep=0pt
     \colorbox{grannysmithapple}{\strut##1}%
   }%
}

\newcommand\HLoff{%
   \xdef\lst@alloverstyle##1{##1}%
}

\newcommand\HLx{%
   \gdef\lst@alloverstyle##1{%
     \fboxrule=0pt
     \fboxsep=0pt
     \colorbox{red1}{\strut##1}%
   }%
}

\newcommand\HLxoff{%
   \xdef\lst@alloverstyle##1{##1}%
}

\newcommand\HLm{%
   \gdef\lst@alloverstyle##1{%
     \fboxrule=0pt
     \fboxsep=0pt
     \colorbox{lightblue}{\strut##1}%
   }%
}

\newcommand\HLmoff{%
   \xdef\lst@alloverstyle##1{##1}%
}

\newcommand\HLmp{%
   \gdef\lst@alloverstyle##1{%
     \fboxrule=0pt
     \fboxsep=0pt
     \colorbox{hlyellow}{\strut##1}%
   }%
}

\newcommand\HLmpoff{%
   \xdef\lst@alloverstyle##1{##1}%
}

\newcommand\HLhl{%
   \gdef\lst@alloverstyle##1{%
     \fboxrule=0pt
     \fboxsep=0pt
     \colorbox{lightgray}{\strut##1}%
   }%
}

\newcommand\HLhloff{%
   \xdef\lst@alloverstyle##1{##1}%
}

%% file: tables/dataset_replacement.tex
\begin{table*}[t]
\centering
\begin{minipage}[t]{0.68\textwidth}
  \centering
  \captionof{table}{Summary of datasets.}
  \label{t:datasets}
  \vspace{-0.1in}
  \small
  \setlength{\tabcolsep}{3.2pt}
  \begin{tabular}{@{}p{12em}@{} 
                  >{\raggedleft\arraybackslash}p{4.5em} |
                  >{\raggedleft\arraybackslash}p{6em} |
                  >{\raggedleft\arraybackslash}p{5.5em} |
                  >{\raggedleft\arraybackslash}p{3.3em} |
                  >{\raggedleft\arraybackslash}p{3.3em} |
                  >{\raggedleft\arraybackslash}p{3.3em}@{}}
      \toprule
      & \multicolumn{2}{c}{Annotated data} 
      & \multirow{2}{*}{\makecell{Suppression\\study data}} 
      & \multicolumn{3}{c}{CodeTracker data} \\
      \cmidrule(lr){2-3} \cmidrule(lr){5-7}
      & Data~A & Data~B & & Variable & Block & Method \\
      \midrule
      Programming language(s) 
      & \multicolumn{2}{l}{\makecell{Python, Java, JavaScript, C\#, C++,\\Go, Ruby, TypeScript, PHP, HTML}} 
      & Python 
      & \multicolumn{3}{c}{Java} \\

      Number of projects 
      & \multicolumn{2}{c}{20} & 8 
      & \multicolumn{3}{c}{10} \\

      Number of source-target pairs 
      & 100 & 100 & 187 
      & 530 & 949 & 526 \\

      Avg. char size of source region 
      & 210.5 & 39.8 & 26.1 
      & 7.8 & 784.6 & 930.4 \\

      Avg. char size of target region 
      & 204.8 & 31.5 & 13.3 
      & 7.7 & 742.2 & 904.8 \\

      Avg. commit distance 
      & 2.1 & 2.0 & -- 
      & 15232.5 & 4626.6 & 3695.0 \\

      Direction of change 
      & \multicolumn{2}{c}{\makecell{50\% forward, 50\% backward}} 
      & 100\% forward 
      & \multicolumn{3}{c}{100\% backward} \\
      \bottomrule
    \end{tabular}
\end{minipage}%
\hfill
\hspace{-0.1in} 
\begin{minipage}[t]{0.3\textwidth}
  \vspace{-0.08in} 
  \centering
  \small
  \[
  \mathit{recall} = \frac{\mathit{overlapping\_chars}}{\mathit{all\_ground\_truth\_chars}}
  \]
  \vspace{.1em}
  \[
  \mathit{precision} = \frac{\mathit{overlapping\_chars}}{\mathit{all\_target\_region\_chars}}
  \]
  \vspace{.1em}
  \[
  \operatorname{F1\text{-}score} = 2 \cdot \frac{\mathit{recall} \cdot \mathit{precision}}{\mathit{recall} + \mathit{precision}}
  \]
  \vspace{.1em}
  \hspace*{1.5em}
  \begin{minipage}{0.9\textwidth}
      \raggedright
      \small
      * \textit{overlapping\_chars} is the number of common characters between the target region and the ground truth.
  \end{minipage}

  \vspace{0.05em}  
  \[
  \mathit{char\_dist} = \left| i - i' \right| + \left| j - j' \right|
  \]

  \vspace{-0.15in}
  \captionof{figure}{Evaluation metrics.}
  \label{fig:metrics}
\end{minipage}
\end{table*}

%% file: tables/ressult_rq1_1.tex
\begin{table*}[t]
  \centering
  \caption{Results on Data A, Data B, and Suppression study data.}
  \vspace{-0.1in}
  \setlength{\tabcolsep}{4.6pt}
  \begin{tabular}{@{}lrrrrrrrrr@{}}
    \toprule
    \multirow{2}{*}{} & \multicolumn{3}{c}{Data A} & \multicolumn{3}{c}{Data B} & \multicolumn{3}{c}{Suppression study data} \\
    \cmidrule(lr){2-4} \cmidrule(lr){5-7} \cmidrule(lr){8-10}
    & diff\textsubscript{line} & diff\textsubscript{word} & \name{}
    & diff\textsubscript{line} & diff\textsubscript{word} & \name{}
    & diff\textsubscript{line} & diff\textsubscript{word} & \name{} \\
    \midrule
    Overlapping     & 89 (89.0\%) & 86 (86.0\%) & \textbf{95 (95.0\%)} 
                    & \textbf{91 (91.0\%)} & 87 (87.0\%) & \textbf{91 (91.0\%)} 
                    & 133 (71.1\%) & 125 (66.8\%) & \textbf{156 (83.4\%)} \\
    Exact matches   & 43 (43.0\%) & 54 (54.0\%) & \textbf{77 (77.0\%)} 
                    & 44 (44.0\%) & 54 (54.0\%) & \textbf{71 (71.0\%)} 
                    & 41 (21.9\%) & 41 (21.9\%) & \textbf{151 (80.7\%)} \\
    Char. dist.     & 147.0 & 71.1 & \textbf{55.2} 
                    & 59.1 & 61.7 & \textbf{18.5} 
                    & 199.4 & 194.1 & \textbf{29.2} \\
    Recall          & 0.890 & 0.849 & \textbf{0.932} 
                    & \textbf{0.906} & 0.859 & 0.892 
                    & 0.709 & 0.666 & \textbf{0.834} \\
    Precision       & 0.656 & 0.741 & \textbf{0.889} 
                    & 0.681 & 0.721 & \textbf{0.883} 
                    & 0.351 & 0.377 & \textbf{0.824} \\
    F1-score        & 0.705 & 0.766 & \textbf{0.898} 
                    & 0.721 & 0.741 & \textbf{0.882} 
                    & 0.412 & 0.433 & \textbf{0.827} \\
    \bottomrule
  \end{tabular}
  \label{tab:result_anno_supp}
\end{table*}

%% file: tables/ressult_rq1_2.tex
\begin{table*}[t]
  \centering
  \caption{Results on CodeTracker data: Variable, Block, and Method.}
  \vspace{-0.1in}
  \setlength{\tabcolsep}{3pt}
  \begin{tabular}{@{}lrrrrrrrrr@{}}
    \toprule
    \multirow{2}{*}{} & \multicolumn{3}{c}{Variable} & \multicolumn{3}{c}{Block} & \multicolumn{3}{c}{Method} \\
    \cmidrule(lr){2-4} \cmidrule(lr){5-7} \cmidrule(lr){8-10}
    & diff\textsubscript{line} & diff\textsubscript{word} & \name{} 
    & diff\textsubscript{line} & diff\textsubscript{word} & \name{}
    & diff\textsubscript{line} & diff\textsubscript{word} & \name{} \\
    \midrule
    Overlapping     & \textbf{432 (81.5\%)} & 363 (68.5\%) & 414 (78.1\%) 
                    & \textbf{890 (93.8\%)} & \textbf{890 (93.8\%)} & 871 (91.8\%) 
                    & \textbf{516 (98.1\%)} & \textbf{516 (98.1\%)} & \textbf{516 (98.1\%)} \\
    Exact matches   & 225 (42.5\%) & 226 (42.6\%) & \textbf{401 (75.7\%)} 
                    & 702 (74.0\%) & 746 (78.6\%) & \textbf{776 (81.8\%)} 
                    & 449 (85.4\%) & 489 (93.0\%) & \textbf{497 (94.5\%)} \\
    Char. dist.     & 643.6 & 882.6 & \textbf{75.8} 
                    & 989.6 & 1287.2 & \textbf{126.4} 
                    & 615.8 & 1522.0 & \textbf{384.3} \\
    Recall          & \textbf{0.815} & 0.685 & 0.781 
                    & \textbf{0.927} & 0.924 & 0.904 
                    & \textbf{0.977} & \textbf{0.977} & \textbf{0.977} \\
    Precision       & 0.449 & 0.455 & \textbf{0.764} 
                    & 0.877 & 0.880 & \textbf{0.900} 
                    & 0.962 & 0.962 & \textbf{0.974} \\
    F1-score        & 0.468 & 0.469 & \textbf{0.767} 
                    & 0.885 & 0.886 & \textbf{0.898} 
                    & 0.962 & 0.962 & \textbf{0.972} \\
    \bottomrule
  \end{tabular}
  \label{tab:result_codetracker_data}
\end{table*}

%% file: references.bib
@misc{post1,
  author = {Stack Overflow User},
  title = {Git: Show all of the various changes to a single line in a specified file over the entire git history},
  year = {2012},
  url = {https://stackoverflow.com/questions/9935379/git-show-all-of-the-various-changes-to-a-single-line-in-a-specified-file-over-t},
  note = {}
}

@misc{post2,
  author = {Stack Overflow User},
  title = {git: grep file in all previous versions},
  year = {2016},
  url = {https://stackoverflow.com/questions/34576699/git-grep-file-in-all-previous-versions?noredirect=1},
  note = {}
}

@misc{post3,
  author = {Stack Overflow User},
  title = {How can I use git log and only output the matching lines?},
  year = {2016},
  url = {https://stackoverflow.com/questions/40936797/how-can-i-use-git-log-and-only-output-the-matching-lines?rq=3},
  note = {}
}

@inproceedings{MSR2006_cochange,
author = {Zimmermann, Thomas and Kim, Sunghun and Zeller, Andreas and Whitehead, E. James},
title = {Mining version archives for co-changed lines},
year = {2006},
isbn = {1595933972},
publisher = {Association for Computing Machinery},
address = {New York, NY, USA},
url = {https://doi.org/10.1145/1137983.1138001},
doi = {10.1145/1137983.1138001},
booktitle = {Proceedings of the 2006 International Workshop on Mining Software Repositories},
pages = {72–75},
numpages = {4},
location = {Shanghai, China},
series = {MSR '06}
}

@INPROCEEDINGS{MSR2007_line_levinshtein_dist,
  author={Canfora, Gerardo and Cerulo, Luigi and Di Penta, Massimiliano},
  booktitle={Fourth International Workshop on Mining Software Repositories (MSR'07:ICSE Workshops 2007)}, 
  title={Identifying Changed Source Code Lines from Version Repositories}, 
  year={2007},
  volume={},
  number={},
  pages={14-14},
  doi={10.1109/MSR.2007.14}}

@inproceedings{MSR2006_element_matching_survey,
author = {Kim, Miryung and Notkin, David},
title = {Program element matching for multi-version program analyses},
year = {2006},
isbn = {1595933972},
publisher = {Association for Computing Machinery},
address = {New York, NY, USA},
url = {https://doi.org/10.1145/1137983.1137999},
doi = {10.1145/1137983.1137999},
booktitle = {Proceedings of the 2006 International Workshop on Mining Software Repositories},
pages = {58–64},
numpages = {7},
location = {Shanghai, China},
series = {MSR '06}
}

@inproceedings{ICSM2004_Meta_differencing,
author = {Maletic, Jonathan I. and Collard, Michael L.},
title = {Supporting Source Code Difference Analysis},
year = {2004},
isbn = {0769522130},
publisher = {IEEE Computer Society},
address = {USA},
booktitle = {Proceedings of the 20th IEEE International Conference on Software Maintenance},
pages = {210–219},
numpages = {10},
series = {ICSM '04}
}

@inproceedings{FSE12_History_slicing,
author = {Servant, Francisco and Jones, James A.},
title = {History slicing: assisting code-evolution tasks},
year = {2012},
isbn = {9781450316149},
publisher = {Association for Computing Machinery},
address = {New York, NY, USA},
url = {https://doi.org/10.1145/2393596.2393646},
doi = {10.1145/2393596.2393646},
booktitle = {Proceedings of the ACM SIGSOFT 20th International Symposium on the Foundations of Software Engineering},
articleno = {43},
numpages = {11},
location = {Cary, North Carolina},
series = {FSE '12}
}

@inproceedings{ICSE17_Fuzzy_history_slicing,
author = {Servant, Francisco and Jones, James A.},
title = {Fuzzy fine-grained code-history analysis},
year = {2017},
isbn = {9781538638682},
publisher = {IEEE Press},
url = {https://doi.org/10.1109/ICSE.2017.74},
doi = {10.1109/ICSE.2017.74},
booktitle = {Proceedings of the 39th International Conference on Software Engineering},
pages = {746–757},
numpages = {12},
location = {Buenos Aires, Argentina},
series = {ICSE '17}
}

@InProceedings{fse25_suppression_study,
  author    = {Huimin Hu and Yingying Wang and Julia Rubin and Michael Pradel},
  title     = {An Empirical Study of Suppressed Static Analysis Warnings},
  booktitle = {FSE},
  year      = {2025},
}

@inproceedings{Codoban2015,
   author = "M. Codoban and S. S. Ragavan and D. Dig and B. Bailey",
   title = "Software history under the lens: A study on why and how developers examine it",
   booktitle = "2015 IEEE International Conference on Software Maintenance and Evolution (ICSME)",
   month = sep,
   year = "2015",
   pages = "1-10"
}

@inproceedings{icse_2021_codeshovel,
  title = {{CodeShovel}: {C}onstructing Method-Level Source Code Histories},
  author = {Felix Grund and Shaiful Alam Chowdhury and Nick Bradley and Braxton Hall and Reid Holmes},
  booktitle = {Proceedings of the International Conference on Software Engineering (ICSE)},
  year = {2021}
}

@inproceedings{10.1145/3540250.3549079,
   author = {Jodavi, Mehran and Tsantalis, Nikolaos},
   title = {Accurate Method and Variable Tracking in Commit History},
   year = {2022},
   isbn = {9781450394130},
   publisher = {Association for Computing Machinery},
   address = {New York, NY, USA},
   url = {https://doi.org/10.1145/3540250.3549079},
   doi = {10.1145/3540250.3549079},
   booktitle = {Proceedings of the 30th ACM Joint European Software Engineering Conference and Symposium on the Foundations of Software Engineering},
   pages = {183–195},
   numpages = {13},
   location = {Singapore, Singapore},
   series = {ESEC/FSE 2022}
}

@article{2020FinerGit,
   author = {Higo, Yoshiki and Hayashi, Shinpei and Kusumoto, Shinji},
   title = {On Tracking Java Methods with Git Mechanisms},
   year = {2020},
   issue_date = {July 2020},
   publisher = {Elsevier Science Inc.},
   address = {USA},
   volume = {165},
   issn = {0164-1212},
   url = {https://doi.org/10.1016/j.jss.2020.110571},
   doi = {10.1016/j.jss.2020.110571},
   journal = {Journal of Systems and Software},
   numpages = {13}
}

@inproceedings{2011Historage,
   author = {Hata, Hideaki and Mizuno, Osamu and Kikuno, Tohru},
   title = {Historage: fine-grained version control system for Java},
   year = {2011},
   isbn = {9781450308489},
   publisher = {Association for Computing Machinery},
   address = {New York, NY, USA},
   url = {https://doi.org/10.1145/2024445.2024463},
   doi = {10.1145/2024445.2024463},
   booktitle = {Proceedings of the 12th International Workshop on Principles of Software Evolution and the 7th Annual ERCIM Workshop on Software Evolution},
   pages = {96–100},
   numpages = {5},
   location = {Szeged, Hungary},
   series = {IWPSE-EVOL '11}
}

@INPROCEEDINGS{2023MergeGen,
  author={Dong, Jinhao and Zhu, Qihao and Sun, Zeyu and Lou, Yiling and Hao, Dan},
  booktitle={2023 38th IEEE/ACM International Conference on Automated Software Engineering (ASE)}, 
  title={Merge Conflict Resolution: Classification or Generation?}, 
  year={2023},
  volume={},
  number={},
  pages={1652-1663},
  doi={10.1109/ASE56229.2023.00155}}

@inproceedings{higo2017generating,
  title={Generating simpler ast edit scripts by considering copy-and-paste},
  author={Higo, Yoshiki and Ohtani, Akio and Kusumoto, Shinji},
  booktitle={2017 32nd IEEE/ACM International Conference on Automated Software Engineering (ASE)},
  pages={532--542},
  year={2017},
  organization={IEEE}
}

@inproceedings{2023HyperDiff,
author = {Le Dilavrec, Quentin and Khelladi, Djamel Eddine and Blouin, Arnaud and J\'{e}z\'{e}quel, Jean-Marc},
title = {HyperDiff: Computing Source Code Diffs at Scale},
year = {2023},
isbn = {9798400703270},
publisher = {Association for Computing Machinery},
address = {New York, NY, USA},
url = {https://doi.org/10.1145/3611643.3616312},
doi = {10.1145/3611643.3616312},
booktitle = {Proceedings of the 31st ACM Joint European Software Engineering Conference and Symposium on the Foundations of Software Engineering},
pages = {288–299},
numpages = {12},
location = {<conf-loc>, <city>San Francisco</city>, <state>CA</state>, <country>USA</country>, </conf-loc>},
series = {ESEC/FSE 2023}
}

@inproceedings{2013LHDiff,
author = {Asaduzzaman, Muhammad and Roy, Chanchal K. and Schneider, Kevin A. and Penta, Massimiliano Di},
title = {LHDiff: A Language-Independent Hybrid Approach for Tracking Source Code Lines},
year = {2013},
isbn = {9780769549811},
publisher = {IEEE Computer Society},
address = {USA},
url = {https://doi.org/10.1109/ICSM.2013.34},
doi = {10.1109/ICSM.2013.34},
booktitle = {Proceedings of the 2013 IEEE International Conference on Software Maintenance},
pages = {230–239},
numpages = {10},
series = {ICSM '13}
}

@inproceedings{canfora2008tracking,
  title={Tracking your changes: A language-independent approach},
  author={Canfora, Gerardo and Cerulo, Luigi and Di Penta, Massimiliano},
  journal={IEEE software},
  volume={26},
  number={1},
  pages={50--57},
  year={2008},
  publisher={IEEE}
}

@inproceedings{2008Trackinglocations,
author = {Reiss, Steven P.},
title = {Tracking source locations},
year = {2008},
isbn = {9781605580791},
publisher = {Association for Computing Machinery},
address = {New York, NY, USA},
url = {https://doi.org/10.1145/1368088.1368091},
doi = {10.1145/1368088.1368091},
booktitle = {Proceedings of the 30th International Conference on Software Engineering},
pages = {11–20},
numpages = {10},
location = {Leipzig, Germany},
series = {ICSE '08}
}

@inproceedings{2019BeyondGumTree,
author = {Matsumoto, Junnosuke and Higo, Yoshiki and Kusumoto, Shinji},
title = {Beyond GumTree: a hybrid approach to generate edit scripts},
year = {2019},
publisher = {IEEE Press},
url = {https://doi.org/10.1109/MSR.2019.00082},
doi = {10.1109/MSR.2019.00082},
booktitle = {Proceedings of the 16th International Conference on Mining Software Repositories},
pages = {550–554},
numpages = {5},
location = {Montreal, Quebec, Canada},
series = {MSR '19}
}

@ARTICLE{AugmentingDiffs2023,
  author={Etemadi, Khashayar and Sharma, Aman and Madeiral, Fernanda and Monperrus, Martin},
  journal={IEEE Transactions on Software Engineering}, 
  title={Augmenting Diffs With Runtime Information}, 
  year={2023},
  volume={49},
  number={11},
  pages={4988-5007},
  doi={10.1109/TSE.2023.3324258}}

@article{DiffSearch2023,
author = {Grazia, Luca Di and Bredl, Paul and Pradel, Michael},
title = {DiffSearch: A Scalable and Precise Search Engine for Code Changes},
year = {2023},
issue_date = {April 2023},
publisher = {IEEE Press},
volume = {49},
number = {4},
issn = {0098-5589},
url = {https://doi.org/10.1109/TSE.2022.3218859},
doi = {10.1109/TSE.2022.3218859},
journal = {IEEE Trans. Softw. Eng.},
month = {apr},
pages = {2366–2380},
numpages = {15}
}

@inproceedings{DIFFBASE2021,
author = {Wu, Xiuheng and Zhu, Chenguang and Li, Yi},
title = {DIFFBASE: a differential factbase for effective software evolution management},
year = {2021},
isbn = {9781450385626},
publisher = {Association for Computing Machinery},
address = {New York, NY, USA},
url = {https://doi.org/10.1145/3468264.3468605},
doi = {10.1145/3468264.3468605},
booktitle = {Proceedings of the 29th ACM Joint Meeting on European Software Engineering Conference and Symposium on the Foundations of Software Engineering},
pages = {503–515},
numpages = {13},
location = {Athens, Greece},
series = {ESEC/FSE 2021}
}

@INPROCEEDINGS{finegrainedchangetypes2016,
  author={Lin, Wei and Chen, Zhifei and Ma, Wanwangying and Chen, Lin and Xu, Lei and Xu, Baowen},
  booktitle={2016 IEEE International Conference on Software Maintenance and Evolution (ICSME)}, 
  title={An Empirical Study on the Characteristics of Python Fine-Grained Source Code Change Types}, 
  year={2016},
  volume={},
  number={},
  pages={188-199},
  doi={10.1109/ICSME.2016.25}}

@inproceedings{Hard-to-answer2010,
author = {LaToza, Thomas D. and Myers, Brad A.},
title = {Hard-to-answer questions about code},
year = {2010},
isbn = {9781450305471},
publisher = {Association for Computing Machinery},
address = {New York, NY, USA},
url = {https://doi.org/10.1145/1937117.1937125},
doi = {10.1145/1937117.1937125},
booktitle = {Evaluation and Usability of Programming Languages and Tools},
articleno = {8},
numpages = {6},
location = {Reno, Nevada},
series = {PLATEAU '10}
}

@inproceedings{LibvDiff2024,
author = {Dong, Chaopeng and Li, Siyuan and Yang, Shouguo and Xiao, Yang and Wang, Yongpan and Li, Hong and Li, Zhi and Sun, Limin},
title = {LibvDiff: Library Version Difference Guided OSS Version Identification in Binaries},
year = {2024},
isbn = {9798400702174},
publisher = {Association for Computing Machinery},
address = {New York, NY, USA},
url = {https://doi.org/10.1145/3597503.3623336},
doi = {10.1145/3597503.3623336},
booktitle = {Proceedings of the IEEE/ACM 46th International Conference on Software Engineering},
articleno = {66},
numpages = {12},
location = {Lisbon, Portugal},
series = {ICSE '24}
}

@INPROCEEDINGS{B2SFinder2019,
  author={Yuan, Zimu and Feng, Muyue and Li, Feng and Ban, Gu and Xiao, Yang and Wang, Shiyang and Tang, Qian and Su, He and Yu, Chendong and Xu, Jiahuan and Piao, Aihua and Xuey, Jingling and Huo, Wei},
  booktitle={2019 34th IEEE/ACM International Conference on Automated Software Engineering (ASE)}, 
  title={B2SFinder: Detecting Open-Source Software Reuse in COTS Software}, 
  year={2019},
  pages={1038-1049},
  doi={10.1109/ASE.2019.00100}
}

@article{similarity-review2019,
author = {Novak, Matija and Joy, Mike and Kermek, Dragutin},
title = {Source-code Similarity Detection and Detection Tools Used in Academia: A Systematic Review},
year = {2019},
issue_date = {September 2019},
publisher = {Association for Computing Machinery},
address = {New York, NY, USA},
volume = {19},
number = {3},
url = {https://doi.org/10.1145/3313290},
doi = {10.1145/3313290},
journal = {ACM Trans. Comput. Educ.},
month = {may},
articleno = {27},
numpages = {37}
}

@article{Mossad_threshold2020,
author = {Devore-McDonald, Breanna and Berger, Emery D.},
title = {Mossad: defeating software plagiarism detection},
year = {2020},
issue_date = {November 2020},
publisher = {Association for Computing Machinery},
address = {New York, NY, USA},
volume = {4},
number = {OOPSLA},
url = {https://doi.org/10.1145/3428206},
doi = {10.1145/3428206},
journal = {Proc. ACM Program. Lang.},
month = {nov},
articleno = {138},
numpages = {28}
}

@inproceedings{buggy_patch_similarity2021,
author = {Tian, Haoye and Liu, Kui and Kabor\'{e}, Abdoul Kader and Koyuncu, Anil and Li, Li and Klein, Jacques and Bissyand\'{e}, Tegawend\'{e} F.},
title = {Evaluating representation learning of code changes for predicting patch correctness in program repair},
year = {2021},
isbn = {9781450367684},
publisher = {Association for Computing Machinery},
address = {New York, NY, USA},
url = {https://doi.org/10.1145/3324884.3416532},
doi = {10.1145/3324884.3416532},
booktitle = {Proceedings of the 35th IEEE/ACM International Conference on Automated Software Engineering},
pages = {981–992},
numpages = {12},
location = {Virtual Event, Australia},
series = {ASE '20}
}

@article{Hasan:TSE:2024:CodeTracker2.0,
   author = {Hasan, Mohammed Tayeeb and Tsantalis, Nikolaos and Alikhanifard, Pouria},
   journal = {IEEE Transactions on Software Engineering},
   title = {Refactoring-aware Block Tracking in Commit History},
   year = {2024},
   pages = {1-20},
   doi = {10.1109/TSE.2024.3484586}
}

@ARTICLE{TSE22RefactoringMiner2.0,
  author={Tsantalis, Nikolaos and Ketkar, Ameya and Dig, Danny},
  journal={IEEE Transactions on Software Engineering}, 
  title={RefactoringMiner 2.0}, 
  year={2022},
  volume={48},
  number={3},
  pages={930-950},
  doi={10.1109/TSE.2020.3007722}
}


%% file: referencesMP.bib
@InProceedings{fse2022_type_study,
  author    = {Luca Di Grazia and Michael Pradel},
  title     = {The Evolution of Type Annotations in Python: An Empirical Study},
  booktitle = {ESEC/FSE},
  year      = {2022},
}
